\documentclass[11pt,letterpaper]{article}

\usepackage{amsfonts,amssymb,graphics,epsfig,verbatim,bm,latexsym,amsmath,url,amsbsy,graphicx,multicol,multirow,natbib,setspace}
\usepackage[top=1 in, bottom=1 in, left=1 in , right=1 in]{geometry}

\newtheorem{theorem}{Theorem}

\newtheorem{lemma}{Lemma}
\newtheorem{example}{Example}

\renewcommand{\thesection}{\arabic{section}}
\renewcommand{\theequation}{\arabic{section}.\arabic{equation}}

\renewcommand{\thelemma}{\arabic{section}.\arabic{lemma}}

\newcommand{\bt}{\beta}

\newcommand{\ld}{\lambda}
\newcommand{\gm}{\gamma}

\newcommand{\eps}[0]{\ensuremath{\varepsilon}}

\newcommand{\lt}{\left}
\newcommand{\rt}{\right}

\newcommand{\sumij}{\sum_{i=1}^n \sum_{j\neq i}}

\newcommand{\ben}{\begin{enumerate}}
\newcommand{\een}{\end{enumerate}}
\newcommand{\bit}{\begin{itemize}}
\newcommand{\eit}{\end{itemize}}

\newcommand{\eq}[1]{\begin{align}#1\end{align}}
\newcommand{\indf}[1]{1\lt\{ #1 \rt\}}
\newcommand{\eqs}[1]{\begin{align*}#1\end{align*}}
\newcommand{\what}[0]{\ensuremath{\widehat}}

\DeclareMathOperator*{\argmax}{arg\,max}

\title{
Exact Computation of Maximum Rank Correlation Estimator\thanks{
  We thank Co-editor, an anonymous referee, Toru Kitagawa and Mike Veall for helpful comments. Shin gratefully acknowledges support from the Social Sciences and Humanities Research Council of Canada (SSHRC-435-2018-0275).
}
}
\author{Youngki Shin and Zvezdomir Todorov}
\author{
Youngki Shin\thanks{
Address: Department of Economics, McMaster University, 1280 Main Street West, Hamilton, Ontario, Canada L8S 4M4. Email: shiny11@mcmaster.ca.} \and
Zvezdomir Todorov\thanks{
Address: Department of Economics, McMaster University, 1280 Main Street West, Hamilton, Ontario, Canada L8S 4M4. Email:  todorovz@mcmaster.ca.}
}
\date{\today}

\begin{document}

\maketitle

\begin{abstract}
        \noindent In this paper we provide a computation algorithm to get a global solution for the maximum rank correlation estimator using the mixed integer programming (MIP) approach. We construct a new constrained optimization problem by transforming all indicator functions into binary parameters to be estimated and show that it is equivalent to the original problem. We also consider an application of the best subset rank prediction and show that the original optimization problem can be reformulated as MIP. We derive the non-asymptotic bound for the tail probability of the predictive performance measure.  We investigate the performance of the MIP algorithm by an empirical example and Monte Carlo simulations.
\medskip


\vspace{0.02in}


\newpage
\end{abstract}

\clearpage

\onehalfspacing

\section{Introduction}

In this paper we provide a computation algorithm to get a global solution for the maximum rank correlation (MRC) estimator using the mixed integer programming (MIP) approach. The new algorithm returns a global solution. The MRC estimator was first proposed by \citet{han1987non} to estimate the generalized regression model:
\eq{
y_i = D \circ F(x'_i\beta,\epsilon), \label{eq:generlaize-regression}
}
where $D:\mathbb{R}\mapsto\mathbb{R}$ is non-degenerate monotonic and $F:\mathbb{R}^2\mapsto\mathbb{R}$ is strictly monotonic in each arguments. The object of interest is the linear index parameter $\bt$.
The model is general enough to include a binary choice model, a censored regression model, and a proportional hazards model as its example. \citet{han1987non} proposed to estimate $\bt$ by maximizing Kendall's rank correlation coefficient:
\eq{
    \what{\bt} = \argmax_{\bt \in \mathcal{B}} \frac{1}{n(n-1)}\sumij \indf{x_i' \bt > x_j' \bt } \indf{y_i > y_j }, \label{eq: obj-mrc1}
}where $\indf{\cdot}$ is an indicator function. He showed the consistency of the MRC estimator and \citet{sherman1993limiting} proved the $\sqrt{n}$-consistency and the asymptotic normality later. The flexible model structure leads to various extensions of the MRC estimator: for example, a quantile index model (\cite{khan2001two}), a generalized panel model (\cite{abrevaya2000rank}), a rank estimation of a nonparametric function (\cite{chen2002rank}), a functional coefficient model (\cite{shin2010local}), a random censoring model (\cite{khan2007partial}), and a partial linear model (\cite{abrevaya2011rank}).

There exist various semiparametric estimators in the class of single-index models (see, for example, the recent work by \citet{ahn2018simple} and the references therein). Compared to them, the MRC estimator has the following advantages. First, it does not require any bandwidth selection since it does not involve any nonparametric estimation components. Second, it can be applied to various models without much modification (see, e.g.,~the references above and \citet*{khan2019inference} for the multinomial models). Finally, it is \emph{point robust} in the sense that it provides a nontrivial identified set that includes the true parameter value when sufficient conditions for point identification are not satisfied (see the discussion in \citet{khan2018discussion} for details).

Implementing the MRC estimator casts some computational challenges in practice, where the grid search method is not feasible. First, the objective function in \eqref{eq: obj-mrc1} is not differentiable in $\beta$ and we cannot apply a gradient-based optimization algorithm. Second, the objective function is not concave. Therefore, any solution found by a numerical algorithm could not be a global solution but a local one. This difficulty is well described in \citet{chay1998estimation}, where they apply Powell's conjugate directions method, the simplex method with multiple starting points, and the piece-wise grid search method repeatedly to achieve a better solution in the empirical application. Even after these repeated searches, we are not sure whether the current solution is the global optimum. Finally, the objective function is the second order U-process and requires $O(n^2)$ computations for a single evaluation. \citet{abrevaya1999computation} shows that the computation order can be reduced to $O(n\log n)$ by adopting the binary search tree structure. However, the fundamental local solution issue still remains.

The contribution of this paper is twofold. First, we propose a new computation algorithm that assures the global solution of the MRC estimator. We achieve this goal by transforming all indicator functions into binary parameters to be estimated along with additional constraints. We show that the proposed mixed integer programming (MIP) problem is equivalent to the original optimization problem. Although MIP is still an NP(non-deterministic polynomial-time)-hard problem (see, e.g.~\citet{wolsey1998integer} and \citet{johnson1978densest} for details), we use a modern mixed integer programming (MIP) solver and confirm that it is feasible to get the solution within a reasonable time budget. The additional advantage of the MIP approach is that it provides us with the gap between the objective function value at the current best solution and the bound of the possible global maximum at any time point of the computation procedure. By this MIP gap, we can measure the quality of the interim solution when the time limit prevents us from waiting for the convergence of the procedure. Second, we consider an application of the best subset rank prediction and analyze the prediction performance. Building on \citet{chen2018best}, we derive a non-asymptotic bound of the tail probability of the predictive performance measure. Since the objective function is defined as a second-order U-process, we develop a new technique to derive the finite sample tail probability bound for higher order U-processes.

We review some related literature. The MIP procedure is recently adopted in various applications in econometrics and statistics. \cite{florios2008exact} show that the maximum score estimator of \citet{manski1975maximum} can be reformulated as an MIP structure. \citet*{bertsimas2016best} consider the best subset selection problem and show that the MIP algorithm outperforms other penalty based methods in terms of achieving sparse solutions with good predictive power. \citet{chen2018best, chen2018exact} investigate the binary prediction problem with variable selection and the instrumental variable quantile regression in the MIP formulation. \citet{kitagawa2018should} apply the MIP procedure when they estimate the personalized optimal welfare policy. Finally, \citet{lee2018factor} develop a MIP computation algorithm to estimate a two-regime regression model when the regime is determined by multi-dimensional factors. To the best of our knowledge, however, this is the first paper in the literature to apply the MIP approach when the objective function is defined as a higher order U-process.

The remainder of the paper is organized as follows. In section 2, we propose the MIP computation algorithm for the maximum rank correlation estimator. We show that the proposed algorithm is equivalent to the original optimization problem and illustrate how it can achieve a feasible solution. In Section 3, we consider the best subset rank prediction problem and derive the non-asymptotic tail probability bound of the performance measure. In section 4, we show the better performance of the proposed MIP algorithm by applying it to the female labor participation data of \cite{mroz1987sensitivity}. Additional numerical evidence is provided through Monte Carlo simulation studies in section 5. We provide some concluding remarks in section 6.

\section{Exact Computation via Mixed Integer Optimization}
In this section we describe the computational challenges of the maximum rank correlation (MRC) estimator and propose a new algorithm to compute a global solution of it. We illustrate the advantage of the new algorithm by investigating a simple numerical example.

We first discuss the computational difficulties of the MRC estimator. Recall that MRC is defined as follows:
\eq{
\what{\bt} = \argmax_{\bt \in \mathcal{B}} \frac{1}{n(n-1)} \sumij \indf{x_i' \bt > x_j' \bt} \indf{y_i > y_j}, \label{eq:obj-mrc}
}where $\mathcal{B}$ is the parameter space of $\bt$ and $\indf{\cdot}$ is an indicator function. Note that the objective function is neither differentiable nor concave in $\bt$. Furthermore, it is defined as a second-order U-process, which requires $O(n^2)$ order of computations for each evaluation of a candidate parameter value.\footnote{\citet{abrevaya1999computation} proposes a nice algorithm that reduces the computation order to $O(n\log n)$ by using the binary search tree. However, it still does not guarantee the global solution.} As a result, we cannot apply any gradient-based optimization algorithm. Researchers usually adopt a simplex-based algorithm such as the Nelder-Meade method in MRC applications. However, it is difficult to get the global solution even  with multiple starting points since the objective function is not globally concave. A grid search algorithm would give more robust solutions but the curse of dimensionality makes it infeasible in most cases when the dimension of $x$ is larger than 2.

In this paper we propose an alternative computational algorithm that is based on the mixed integer programming (MIP) procedure. Let $x_{ij} := x_i - x_j$ be a pairwise difference of $x_i$ and $x_j$. Let $\eps$ be a small positive number, e.g.\ $\eps=10^{-6}$, to denote an effective zero. Consider the following mixed integer programming problem: for $i,j =1,\ldots, n$ and $i \neq j$
\eq{
&\lt(\what{\bt}, \lt\{\what{d}_{ij}\rt\}\rt) = \argmax_{\bt, \{d_{ij}\}} \frac{1}{n(n-1)}\sumij d_{ij} \indf{y_i > y_j} \label{eq:obj-mio}\\
&\mbox{subject to} \notag \\
& \hskip60pt \bt \in \mathcal{B}  \\
& \hskip60pt (d_{ij} - 1) M_{ij} < x_{ij}'\bt\le d_{ij} M_{ij} \label{eq:const2} \\
& \hskip60pt d_{ij} \in \{0,1\} \label{eq:const3}
}where
$
M_{ij} = \max_{\bt \in \mathcal{B}} \lt\vert x_{ij}' \bt \rt\vert + \eps
$.
\footnote{Note that $M_{ij}$ is not a user-chosen turning parameter as it is the empirical bound of $\vert x_{ij}'\beta\vert$ determined by the parameter space and the data.}
Since the objective function in \eqref{eq:obj-mio} is the linear function of the binary variables $\{d_{ij}\}$, the formulation becomes a mixed integer linear programming problem. We check the equivalence between the original problem in \eqref{eq:obj-mrc} and the MIP problem in \eqref{eq:obj-mio}--\eqref{eq:const3}. Consider that the MIP problem chooses $\what{d}_{ij}=1$ for some $i,j$. Then, the constraint \eqref{eq:const2} implies that the estimate for $\what{\bt}$ should satisfy $0<x_{ij}'\what{\bt} \le M_{ij}$, which is equivalent to $x_i\what{\bt} > x_j\what{\bt}$ for a large enough $M_{ij}$. Similarly, $\what{d}_{ij}=0$ is equivalent to $x_i\what{\bt} \le x_j\what{\bt}$. In sum, the constraint forces ${d}_{ij}=1\{x_{ij}'\bt >0\} = 1\{x_i'\bt > x_j'\bt\}$ given any $\bt \in \mathcal{B}$. Therefore, we can compute the global solution $\what{\bt}$ for \eqref{eq:obj-mrc} by solving the equivalent MIP problem in \eqref{eq:obj-mio}--\eqref{eq:const3}. These two optimization problems give us the same numerical results but the MIP procedure has a clear computational advantage over the original problem, which is illustrated below.

\begin{figure}[t]
\caption{Binary Tree Representation of $\{d_{ij}\}$} \label{fig:binary-tree}
\centerline{\includegraphics[scale=.9]{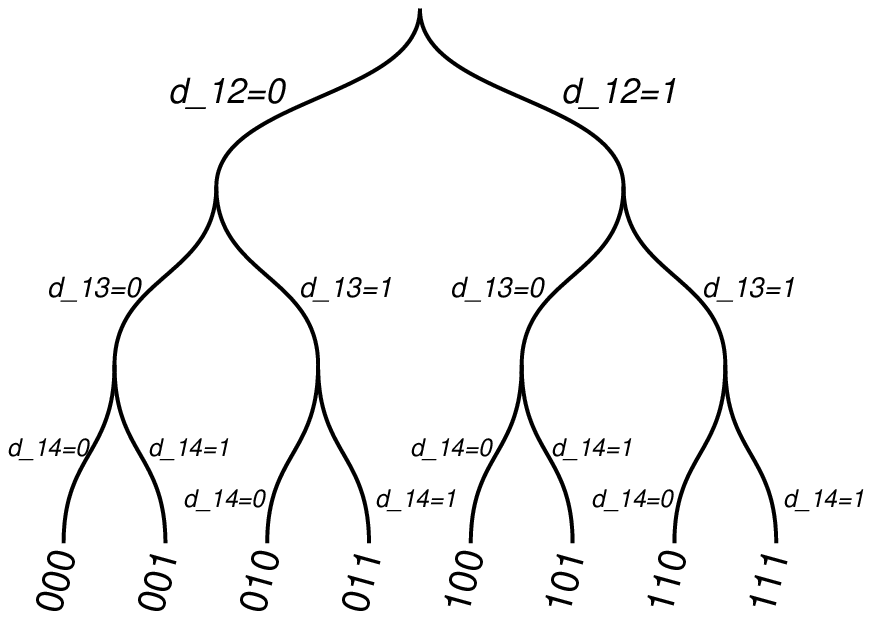}}

  \footnotesize
  Note: The triplet at the bottom of the decision tree denotes a possible choice for $(d_{12},d_{13},d_{14})$. For example, $010$ means $(d_{12},d_{13},d_{14})=(0,1,0)$.

\end{figure}

Modern numerical solvers such as CPLEX and Gurobi make it possible to solve a large scale MIP problem by adopting branch-and-bound type approaches. We provide a heuristic explanation of how a vanilla branch-and-bound algorithm reduces the computational burden followed by a numerical example.
Consider a binary tree representation for all possible values of $\{d_{ij}\}$  (for example, see  Figure \ref{fig:binary-tree}). A bottom node of the tree represents a different possible solution for $\{d_{ij}\}$, and $\bt$ can be easily solved by the linear programming procedure since $d_{ij}$ is fixed there. However, we have $2^{n(n-1)}$ bottom nodes in total and the brute force approach is still infeasible with a standard sample size. The branch-and-bound approach help eliminate a partition of the final nodes systematically.
Suppose that we are in a node located in the middle of the tree, where only a part of $\{d_{ij}\}$ is  fixed. Let $U^*$ be the current best objective function value.\footnote{An initial solution can be achieved from the linear programming problem at any bottom node of the tree.} Now we solve a subproblem after relaxing all $\{d_{ij}\}$ that are not fixed by parent nodes into \emph{continuous} variables on the interval $[0,1]$. This relaxed subproblem can be solved easily by the linear programming procedure since it does not contain integer parameters anymore. There are two cases where we can reduce the computational burden. First, the objective function value of the relaxed problem, say $U^*_R$, is less than or equal to $U^*$. Since the objective function value of the original subproblem is always worse than that of the relaxed subproblem, we cannot achieve a better result than $U^*$ by solving any bottom nodes below the current node. Thus, we can drop all of them from the computation list. Second, $U^*_R > U^*$ and the solution of the relaxed problem satisfies the binary restriction for $\{d_{ij}\}$. This solution coincides with that of the original subproblem. Then, we update $U^* = U^*_R$ and can drop all bottom nodes below it from the computation list. While moving on to a child node, we solve a relaxed subproblem repeatedly and drop a partition of the bottom nodes from our computation list.

We provide a simple numerical example to illustrate how the branch-and-bound algorithm works.
\begin{example}\label{ex:bab}
Consider a sample of $\{(y_i, x_{1i}, x_{2i})\}_{i=1}^4=\{(1,0,2), (0,1,0), (0,1,1), (0,0.5,2)\}$. We normalize $\bt_1=1$ and set the parameter space for $\bt_2$ as $[-5,5]$. There are only three paired observations that satisfy the condition $\{y_i>y_j\}$ and the MIP problem becomes
\eqs{
& \argmax_{\bt_2, d_{12}, d_{13}, d_{14}} \frac{1}{12} \left( d_{12} + d_{13} + d_{14}  \right)\\
& \hskip10pt \mbox{subject to} \notag \\
& \hskip30pt \bt_2 \in [-5,5]  \\
& \hskip30pt (d_{12} - 1) \cdot 11 < -1 + 2\bt_2 \le d_{12}\cdot 11 \\
& \hskip30pt (d_{13} - 1) \cdot 6 < -1 + \bt_2 \le d_{13}\cdot 6 \\
& \hskip30pt (d_{14} - 1) \cdot 1 < -0.5  \le d_{14}\cdot 1 \\
& \hskip30pt d_{12}, d_{13}, d_{14} \in \{0,1\}.
}
Figure \ref{fig:binary-tree} shows the binary tree representation and the brute force approach requires solving 8 linear programming problems at the bottom nodes.  We set $U^* = - \infty$ and solve the first relaxed subproblem at the child node of $d_{12}=1$ (the first right branch in Figure \ref{fig:binary-tree}). The solution for this relaxed subproblem is $(\bt_2, d_{13}, d_{14})=(5,1,0)$ with the objective function value $Q^*_R=2/12$. Since $U^*_R>U^*$ and $(d_{13}, d_{14})$ satisfies the binary restriction, we update $U^*=2/12$ and drop all the nodes below $d_{12}=1$. We next look at the relaxed subproblem at $d_{12}=0$ (the first left branch in Figure \ref{fig:binary-tree}). A solution is $(\bt_2, d_{13}, d_{14})=(1/2, 11/12, 0)$ with the objective function value $U^*_R=23/144$. Since $U^*_R < U^*$, we can drop all the nodes below $d_{12}=0$. Recall that any objective function value from the bottom nodes under $d_{12}=0$ cannot be larger than $23/144$. Therefore, we achieve the solution by solving only two linear programming problems out of the total eight problems.
\end{example}

Finally, we have some remarks on the implementation of the MIP procedure in \eqref{eq:obj-mio}--\eqref{eq:const3}. First, $M_{ij}$ can be computed by solving a separate linear programming problem and saving those values. Alternatively, we can also set a big number $M_{max}$ for all $i,j$, which is large enough to cover the absolute bound of the index $ x_{ij}'\bt$. The second order U-process property requires $O(n^2)$ computations for getting each $M_{ij}$ and it is usually faster to impose a constant number $M_{max}$ for all $M_{ij}$ than to solve the linear programming problem for each $i.j$ in our simulation studies. Second, it is well-known in the numerical optimization literature that any strict inequality should be switched into a weak inequality with some numerical precision bound. Thus, we change the second constraint in \eqref{eq:const2} into
\eqs{
(d_{ij} - 1) M_{ij} + \eps \le x_{ij}'\bt\le d_{ij} M_{ij}.
}
Third, when there are many tied observations in the dependent variable, we can reduce computation cost substantially by vectorizing paired observations and dropping the tied pairs as we have observed in Example \ref{ex:bab}.

%
%
\section{Best Subset Rank Prediction}\label{sec:theory}
In this section we consider the application of a rank prediction problem. The goal is to find a linear index model that gives the best rank prediction of $y$ given $x$. The dimension of $x$ is potentially large and the model selection turns out to be the selection of the best predictors, i.e.~the best subset. We propose an $\ell_0$-constraint maximum rank correlation estimation procedure and show that the MIP method in \eqref{eq:obj-mrc}--\eqref{eq:const3} can be immediately extended to this estimation problem. Building on \citet{chen2018best}, we also provide the non-asymptotic bound of the rank prediction error. This bound implies that the dimension of $x$ can grow exponentially fast if the best subset size grows slowly enough, e.g.~at a polynomial rate.

Suppose that we have a training set of $\{(y_i,x'_i): i=1,\ldots,n\}$, where $y$ can be either discrete or continuous.
We want to learn the rank prediction rule for $y$ as well as to select the best $s$ predictors among $x$'s. Let $x=(x_1,x_{-1}')$ be $(p+1)$ covariates and we know that $x_1$ should be included in the predictor set. Let $\Vert \cdot \Vert_0$ be the $\ell_0$-norm, i.e.\ $\Vert \bt \Vert_0$ is the number of non-zero elements of the vector $\bt$. For any $k\neq l$, we propose the following prediction rule:
\eqs{
    R_{\bt}(x_k,x_l) = 1\{ x_{1,k} + x_{-1,k}'\bt > x_{1,l} + x_{-1,l}'\bt \},
}where $R_{\bt}(x_k,x_l)=1$ implies that $y_k$ is predicted to be larger than $y_l$.
When we are given the whole prediction set $\{x_l: l=1\ldots, n_p\}$, the rank of $y_k$ is predicted by $\sum_{l=1}^{n_p} R_\bt(x_k,x_l)$. Let $F$ be the joint distribution of $(Y,X)$ and $Q:=P\times P$ be the product measure of $P$. Then, we choose the prediction rule as a sample analogue of
 \eqs{
    S(\bt) := Q\lt[1\{y_k > y_l\} = R_{\bt}(x_k,x_l)\rt].
 }
 Recall that we also want to select the best $s$ predictors out of the total $p$ covariates of $x_{-1}$. Therefore, the prediction rule composed of the best $s$ predictors can be achieved by solving the the following $\ell_0$-constraint optimization problem:
\eq{
    \max_{\bt \in \mathcal{B}_s} S_n(\bt), \label{eq:max_problem}
}where $\mathcal{B}_s : = \{\bt \in R^p: \Vert \bt \Vert_0  \le s \}$ and
\eq{
    S_n(\bt) = \frac{2}{n(n-1)} \sum_{i =1}^n \sum_{j > i} 1\{ 1\{y_i> y_j \} = R_{\bt}(x_i,x_j) \}.\label{eq:def-of-S_n}
}

We evaluate the performance of the predictor by the following measure:
\eqs{
    U_n := S^*_{s} - S(\what{\bt}),
}where $S^*_{s}:=\sup_{\bt \in \mathcal{B}_s} S(\bt)$ and $\what{\bt}$ is the solution of the constraint maximization problem defined in \eqref{eq:max_problem}--\eqref{eq:def-of-S_n} above. Note that $U_n \ge 0$ by the definition of $S_n^*$ and that a good prediction rule results in a small value of $S_n$ with a high probability. In the next theorem, we provide a non-asymptotic bound of $U_n$. Let $a \vee b := \max\{a,b\}$ and $r_n:= s\ln(p\vee n) \vee 1$.
\begin{theorem}\label{thm-main}
Suppose that $s \ge 1$. For any $\sigma>0$, there exists a universal constant $D_{\sigma}$ such that
\eq{
    \Pr \lt(  U_n > 4\sqrt{\frac{D_{\sigma}r_n}{n}} \rt) \le \exp(-2 \sigma r_n) \label{eq:thm1-main}
}provided that
\eq{
    &(12s + 12) \ln (D_{\sigma} r_n) \le r_n + (24s+24) \ln2, \label{eq:thm1-con1}\\
    &\lt(8s + \frac{17}{2} \rt) \ln (D_{\sigma} r_n) + (16s + 16)(9 \ln 2 +1) \le r_n. \label{eq:thm1-con2}
}
\end{theorem}
Theorem \ref{thm-main} shows that the tail probability of $U_n$ decreases exponentially in $r_n$. The probability bound in \eqref{eq:thm1-main} is non-asymptotic and holds for every $n$ if two inequality conditions \eqref{eq:thm1-con1}--\eqref{eq:thm1-con2} hold. Compared to the non-asymptotic bound of the best subset selection in \citet{chen2018best}, Theorem \ref{thm-main} requires an additional condition \eqref{eq:thm1-con2} to bound the second order degenerate U-process. However, focusing on the leading terms, we confirm that both conditions hold if
\eq{
12(\ln s + \ln D_{\sigma} + \ln\lt(\ln(p \vee n)\rt) \le \frac{1}{2} \ln (p \vee n ).
}
Note that Theorem \ref{thm-main} implies that $E(U_n) = O(n^{-1/2}\sqrt{s\ln(p\vee n)}) = o(1)$ if $s\ln(p\vee n)=o(n)$. Therefore, the best subset rank prediction performs well even when $p$ grows exponentially provided that $s$ increases slowly, e.g.\ at a polynomial rate.

We finish this section by formulating the $\ell_0$-constraint optimization problem as an MIP problem. Let $x_{-1,ij}:=x_{-1,i} - x_{-1,j}$ as before. For $i,j=1,\ldots,n$, $i\neq j$, $h=1,\ldots,p$, we consider the following constraint MIP problem:
\eq{
& \lt(\what{\bt}, \lt\{\what{d}_{ij}\rt\}, \lt\{\what{e}_{h}\rt\}\rt) = \argmax_{\bt, \{d_{ij}\}, \lt\{{e}_{h}\rt\}} \frac{2}{n(n-1)} \sum_{i =1}^n \sum_{j > i}  \Big[ (1-\indf{y_i > y_j}) + (2\cdot \indf{y_i > y_j}-1) \cdot d_{ij} \Big]  \label{eq:obj-predict}\\
&\mbox{subject to} \notag \\
& \hskip60pt (d_{ij} - 1) M_{ij} < (x_{1,i} - x_{1,j}) + x_{-1,ij}'\bt \le d_{ij} M_{ij} \label{eq:const1-predict} \\
& \hskip60pt e_h \underline{\bt}_h \le \bt_h \le e_h \overline{\bt}_h \label{eq:const2-predict} \\
& \hskip60pt \sum_{h=1}^p e_{h} \le s \label{eq:const3-predict} \\
& \hskip60pt d_{ij} \in \{0,1\} \label{eq:const4-predict} \\
& \hskip60pt e_{h} \in \{0,1\}, h \in \{1,\ldots,p\} \label{eq:const5-predict}
}where $\underline{\bt}_h$ and $\overline{\bt}_h$ are the lower bound and the upper bound of $\bt_h$, respectively. The constraint MIP problem in \eqref{eq:obj-predict}--\eqref{eq:const5-predict} is equivalent to the original constraint optimization problem. The objective function in \eqref{eq:obj-predict} is numerically same with $S_n$ since $d_{ij}$ is identical to $R_{\bt}$ for each $\bt$. Furthermore, the constraint \eqref{eq:const2-predict} makes $\bt_h=0$ whenever $e_h=0$. Thus, the $\ell_0$-norm constraint $\Vert \bt \Vert_0 \le s$ is achieved by the constraints \eqref{eq:const2-predict}, \eqref{eq:const3-predict} and \eqref{eq:const5-predict}.
Note that the objective function can be also written in the familiar rank correlation form:
\eqs{
    \frac{2}{n(n-1)} \sum_{i=1}^n \sum_{j>i} \Big[ 1(y_i>y_j) d_{ij} + 1(y_i\le y_j) (1-d_{ij}) \Big],
}which is equivalent to \eqref{eq:obj-predict}.

%
%
\section{Empirical Illustration}
In this section we illustrate the advantage of the MIP procedure in an empirical application. We revisit the female labor force participation application in \cite{mroz1987sensitivity} and estimate the binary choice model using the generalized regression model in \eqref{eq:generlaize-regression}. Specifically, the composite functions are defined as $F(x'\bt,\eps):=x'\bt+\eps$ and $D(A):=1\{A>0\}$ so that it becomes a semiparametric binary choice model:
\eqs{
  y_i = 1\{x_i'\bt + \eps_i > 0\},
}where the distribution of $\eps_i$ is not specified. The parameter of interest is $\beta$ and we estimate it using the maximum rank correlation estimator defined in \eqref{eq:obj-mrc}. The outcome variable, $y_i$, is 1 if she participated in the labor force and 0, otherwise. We choose the following seven variables from the data for the covariate $x_i$: the number of kids less than 6-year-old ($kidslt6$), the number of kids aged between 6 and 18 ($kidge6$), years of education ($educ$), family income minus her income ($nwifeinc$) in \$1,000, years of experience ($exper$), experience squared ($expersq$), and age ($age$). We randomly draw 100 observations out of 753 for this computation exercise. Table \ref{tb: summary} reports summary statistics of both samples and we confirm that there is not much difference in terms of the mean and the standard deviation of each variable. We normalize the coefficient of $kidslt6$ to be -1. Note that the grid search method is infeasible given the sample size and the number of regressors in this application.

\begin{table}[thp]
\begin{center}
\caption{Summary Statistics} \label{tb: summary}
\begin{tabular}{lrrrr}
  \hline
Variable Names& Mean & Std. Div. & Mean & Std. Div. \\
  \hline
            & \multicolumn{2}{c}{\underline{Subsample}} & \multicolumn{2}{c}{\underline{Original Sample}} \\
Labor Participation & 0.55 & 0.50 & 0.57 & 0.50 \\
\\
  kidslt6   & 0.22 & 0.52  & 0.24 & 0.52 \\
  kidsge6   & 1.54 & 1.31  & 1.35 & 1.32 \\
  educ      & 11.74 & 2.12 & 12.29 & 2.28 \\
  nwifeinc  & 19.66 & 10.64 & 20.13 & 11.63 \\
  exper     & 10.83 & 8.30 & 10.63 & 8.07 \\
  age       & 42.92 & 8.15  & 42.54 & 8.07 \\
\hline
Sample Size  & \multicolumn{2}{c}{100}& \multicolumn{2}{c}{753}\\
\hline
\end{tabular}
\end{center}
\footnotesize
\renewcommand{\baselineskip}{11pt}
\textbf{Note:} The data set is from \citet{mroz1987sensitivity}. The original sample was collected from the Panel Studies of Income Dynamics in 1975. The variable names are explained in the main text.
\end{table}

Table \ref{tb: application} summarizes the estimation results. First, we estimate the model using the mixed integer programming procedure (MIP) with a time budget of 600 seconds. To compare its performance with the existing methods, we also estimate it using the following five methods: the Nelder-Mead simplex method with an initial value from OLS (Nelder-Mead 1), the Nelder-Mead method with multiple initial values until the time budget of 600 seconds is reached (Nelder-Mead 2), the iterative grid search method (Iter-Grid), the simulated annealing method (SANN), and the Markov Chain Monte Carlo (MCMC) method in \citet{chernozhukov2003mcmc}. The parameter space was set to be $\mathcal{B}=[-10,10]^6$. The random starting points of Nelder-Mead 2 was generated from the uniform distribution on $\mathcal{B}$. We use the 2,001 equi-spaced grid points for each parameter for Iter-Grid. The Nelder-Mead method has been adopted in the applications of the MRC estimator, where the grid search is infeasible (for example, see \citet{cavanagh1998rank}, \citet{abrevaya2003pairwise}, \citet{khan2009inference}). A more sophisticated version of the iterative grid search method is introduced by \cite{wang2007note} and it is adopted in \citet*{fan2020rank} for their simulation studies with multi-dimensional regressors.


The estimation result in Table \ref{tb: application} reveals several advantages of MIP over the existing alternative algorithms.
First, MIP achieves the best objective function value among the candidate estimation methods within a reasonable time budget.
Second, some estimates of $\hat{\bt}$ by alternative algorithms are qualitatively different from the solution of MIP. The coefficient estimate of $kidsge6$ by Nelder-Mead 1 shows the opposite direction. The estimates of $educ$ by alternative algorithms show much higher effects than MIP. MCMC shows the closest result although it is still suboptimal.
Third, the Nelder-Mead algorithm with the multiple staring point for 600 seconds does not improve the result. In fact, the objective function value of Nelder-Mead 2 is lower than that of Nelder-Mead 1 which uses only one starting point of the OLS estimate. Finally, Figure \ref{fig:obj-values} shows how difficult the optimization problem is. We plot the empirical objective function values over the convex combinations of two $\beta$ estimates of MIP and Nelder-Mead 1. We can confirm that the objective function is not concave and that there exist many local maxima even between these two estimates.

\begin{table}[thp]
\begin{center}
\caption{Female Labor Participation} \label{tb: application}
\resizebox{\textwidth}{!}{
\begin{tabular}{lrrrrrrrrr}
  \hline
  Method 		& Obj. 	& Time (sec.) 	& kidslt6 & kidsge6 & educ 	 & nwifeinc & exper & expersq & age \\
  \hline
  MIO 			& 0.2140 & 600.33 		& -1.0000 & -0.1523 & 0.0775 & -0.0066 & 0.0480 & 0.0008 & -0.0696 \\
  Nelder-Mead 1 & 0.2087 & 0.27 		& -1.0000 & 0.0385 	& 0.2812 & -0.0147 & 0.2061 & -0.0028 & -0.0533 \\
  Nelder-Mead 2 & 0.2026 & 609.39 		& -1.0000 & -1.3480 & 9.4946 & -0.9316 & 8.9737 & -0.1376 & -2.0547 \\
  Iter-Grid 	& 0.1989 & 4.62 		& -1.0000 & -0.3800 & 2.9300 & -0.0400 & 1.5500 & -0.0100 & -0.2200 \\
  SANN 			& 0.2018 & 6.43 		& -1.0000 & 1.6951 	& 2.1155 & -0.4312 & -0.2469 & 0.3394 & -0.4344 \\
  MCMC 			& 0.2129 & 2.36 		& -1.0000 & -0.1342 & 0.1522 & -0.0077 & 0.0816 & 0.0006 & -0.0788 \\
  \hline
\end{tabular}}
\end{center}
\footnotesize
\renewcommand{\baselineskip}{11pt}
\textbf{Note:} MIP denotes the mixed integer programming method. Nelder-Mead 1 and 2 denote the Nelder-Mead simplex methods with an initial value from OLS and multiple random initial values given the time budget of 600 seconds. Iter-Grid denotes the iterative grid search method with an initial value from OLS. SANN denotes the simulated annealing method. MCMC denotes the Markov Chain Monte Carlo method in \citet{chernozhukov2003mcmc}. The unit of computation time is seconds.
\end{table}

\begin{figure}[thp]
\begin{center}
\caption{Objective Function Values}\label{fig:obj-values}
\includegraphics[scale=0.6]{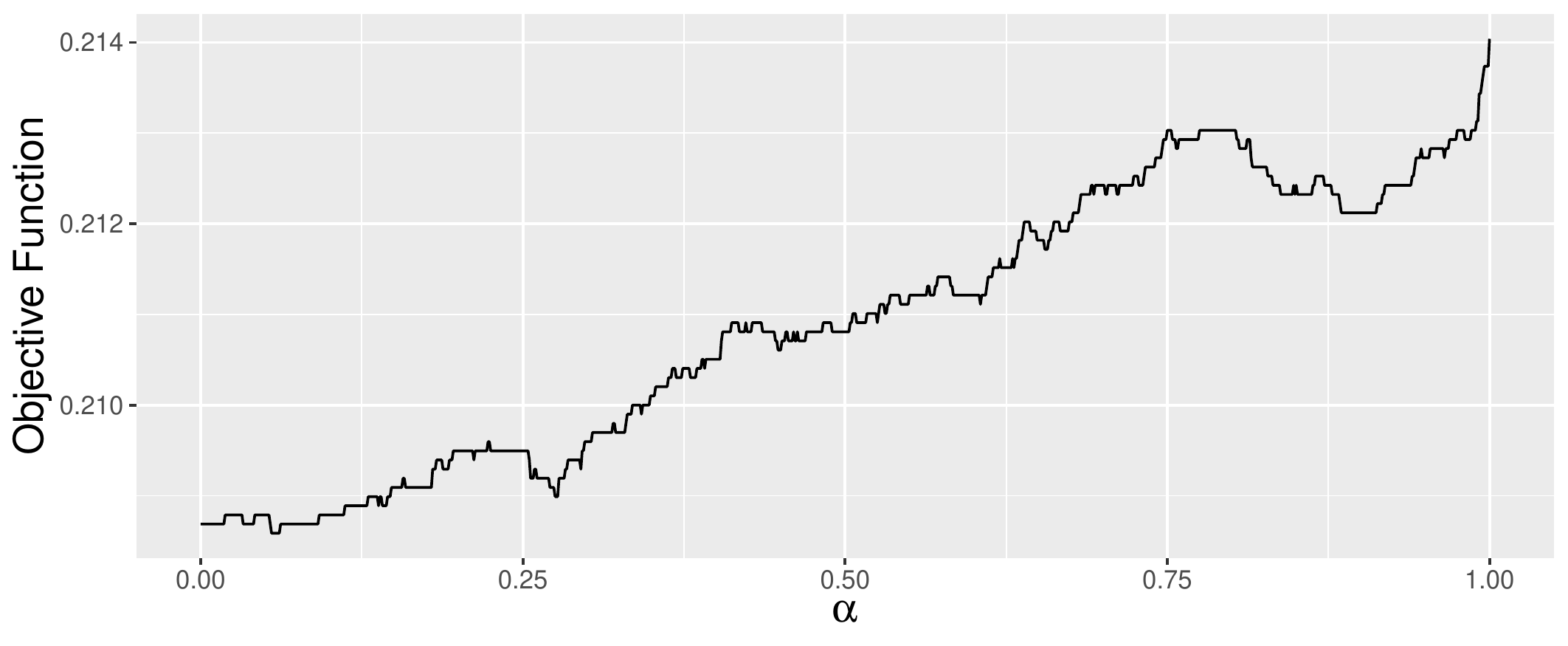}
\end{center}
\footnotesize
\renewcommand{\baselineskip}{11pt}
\textbf{Note:} The empirical objective function values are plotted over the convex combinations of two $\beta$ estimates of MIP and Nelder-Mead 1: $\hat{\beta}_{\alpha} = \alpha \cdot \hat{\beta}_{MIP} + (1-\alpha) \cdot \hat{\beta}_{NM1}$ for $\alpha \in [0,1]$, where $\hat{\beta}_{MIP}$ and $\hat{\beta}_{NM1}$ are MIP and Nelder-Mead 1 estimates, respectively.
\end{figure}

In sum, inference based on inferior local solutions could lead researchers to imprecise or incorrect conclusions in practice although the theoretical properties of the MRC estimator are robust.

%
%
%

\section{Monte Carlo Simulations}
In this section we investigate the performance of the proposed MIP algorithm for the MRC estimator via Monte Carlo simulation studies.
We focus on the achieved objective function value and the computation time in this section.
All simulations are carried out on a computer equipped with AMD Ryzen Threadripper 1950X 16-Core processor and 64 Gigabytes of RAM.

\begin{figure}[hp]
\caption{Loss/Tie/Win Ratio (Binary)} \label{fg:win-bin}
\centering
\begin{tabular}{cc}
$\underline{n=50}$ & $\underline{n=100}$ \\
\includegraphics[scale=0.3]{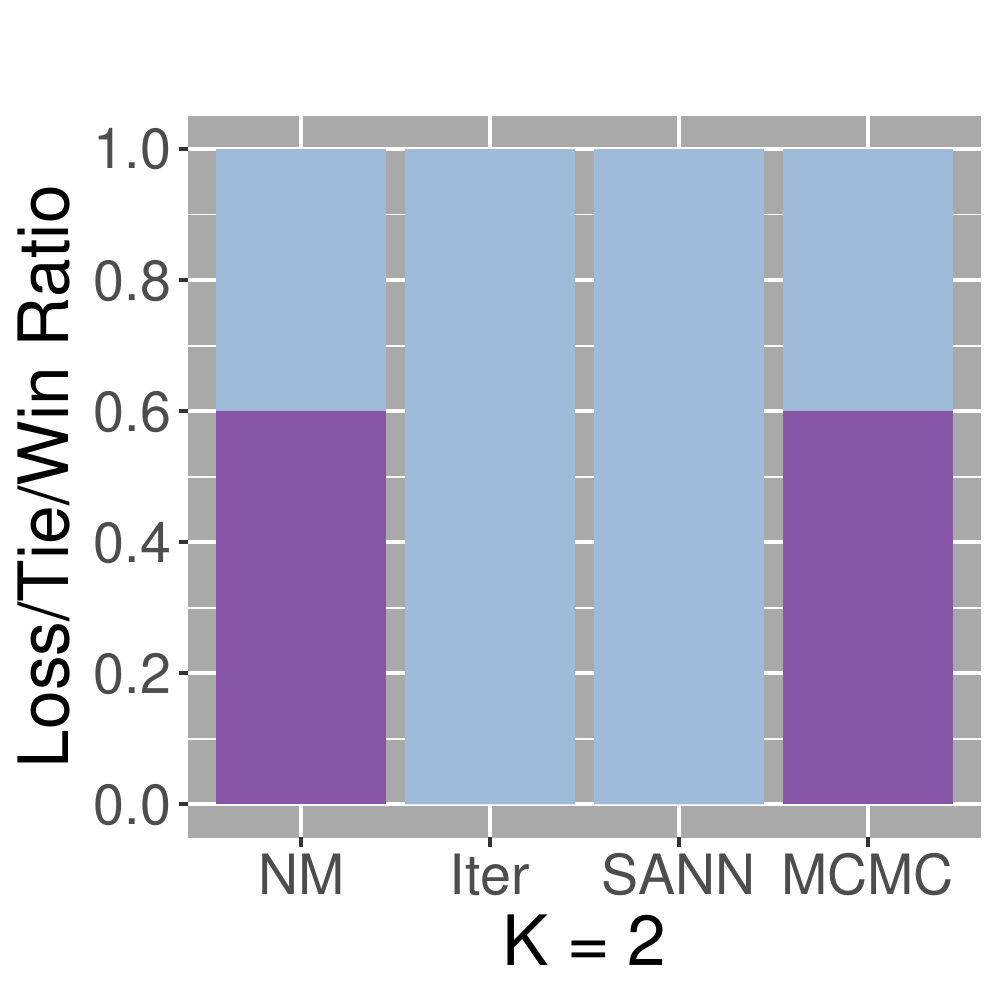}
\includegraphics[scale=0.3]{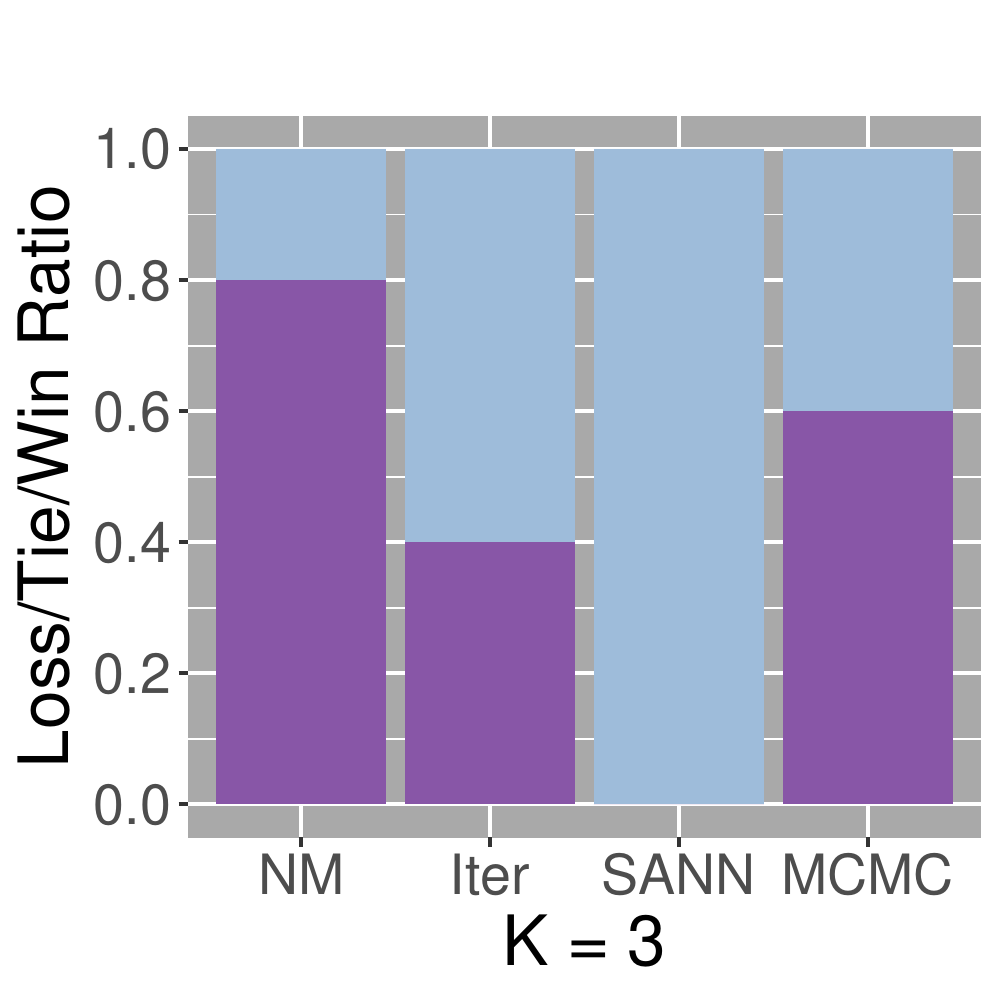} &
\includegraphics[scale=0.3]{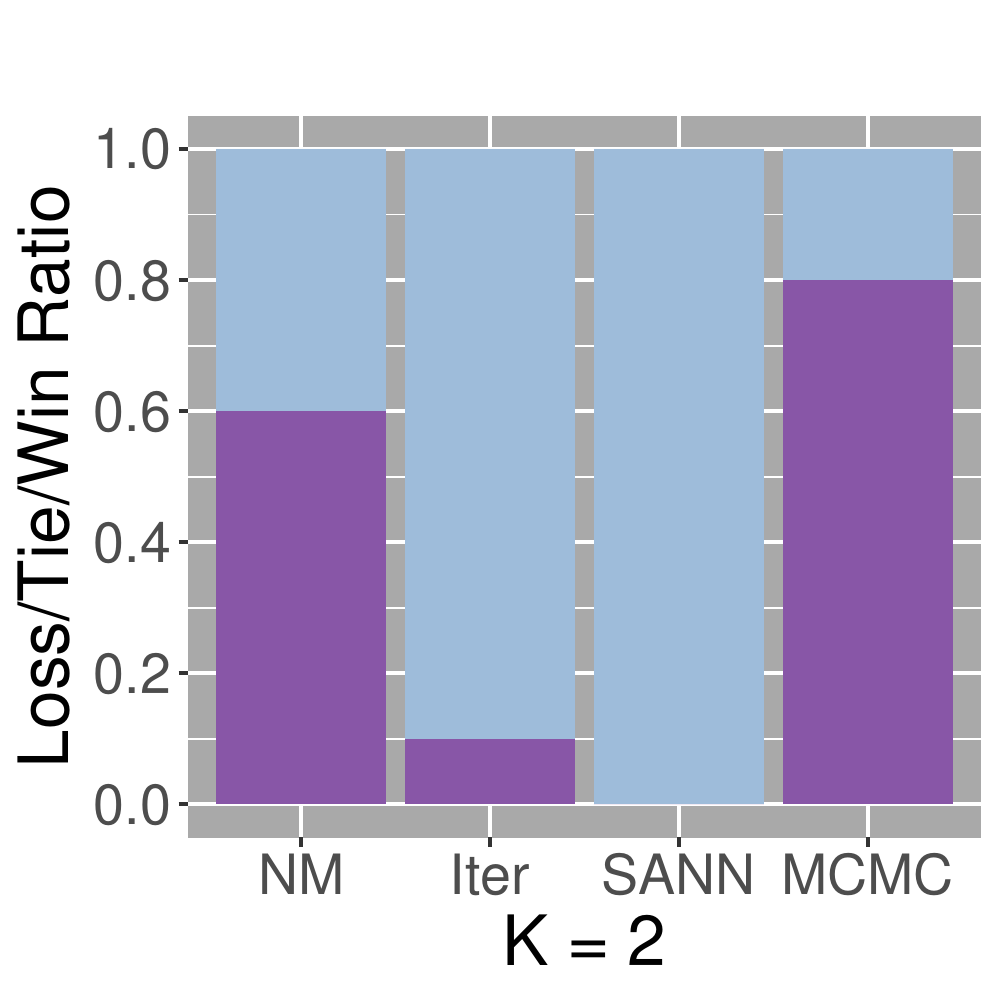}
\includegraphics[scale=0.3]{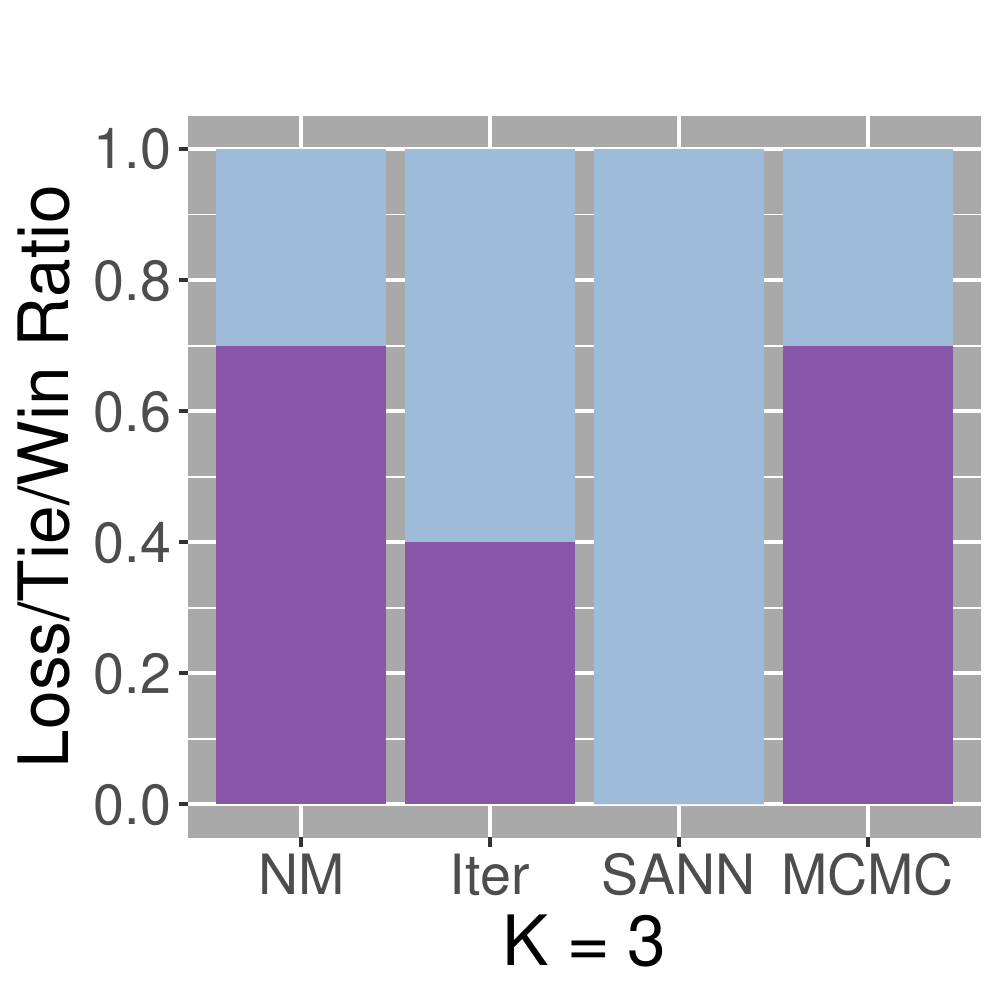}\\
$\underline{n=200}$ & $\underline{n=400}$ \\
\includegraphics[scale=0.3]{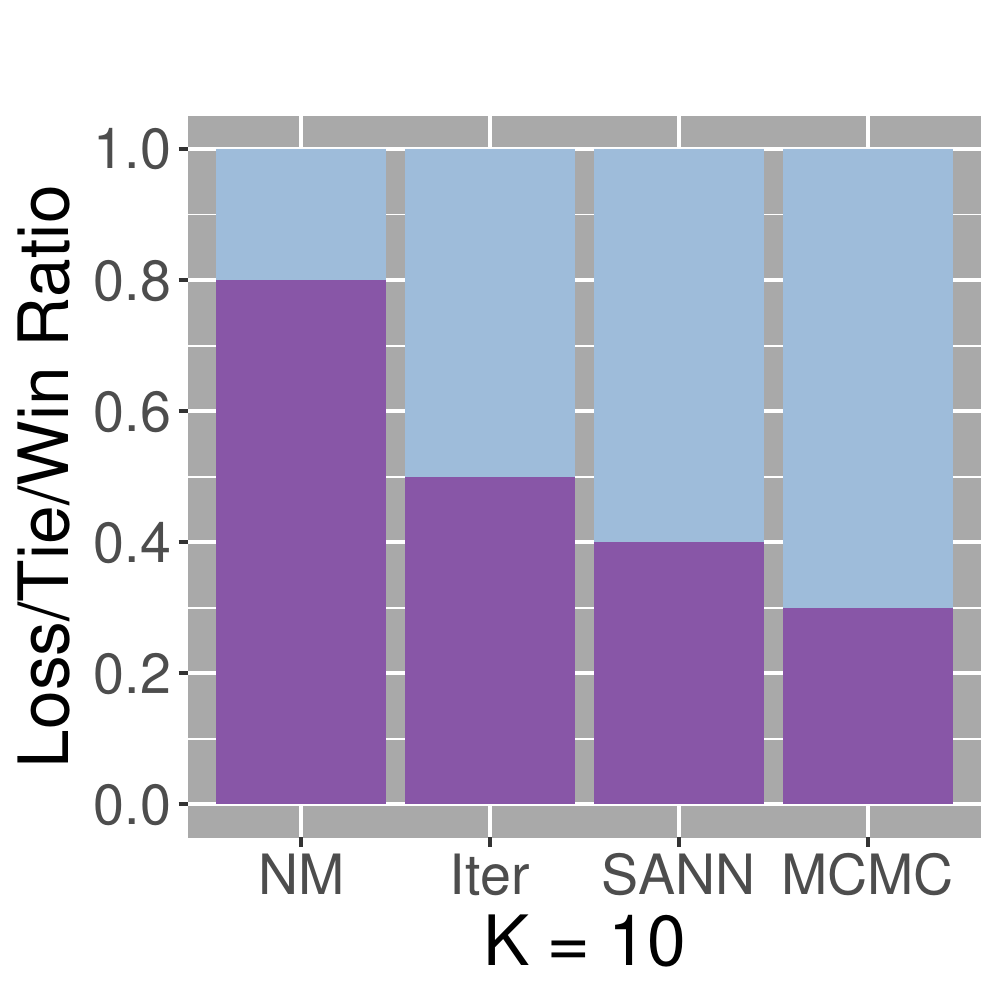}
\includegraphics[scale=0.3]{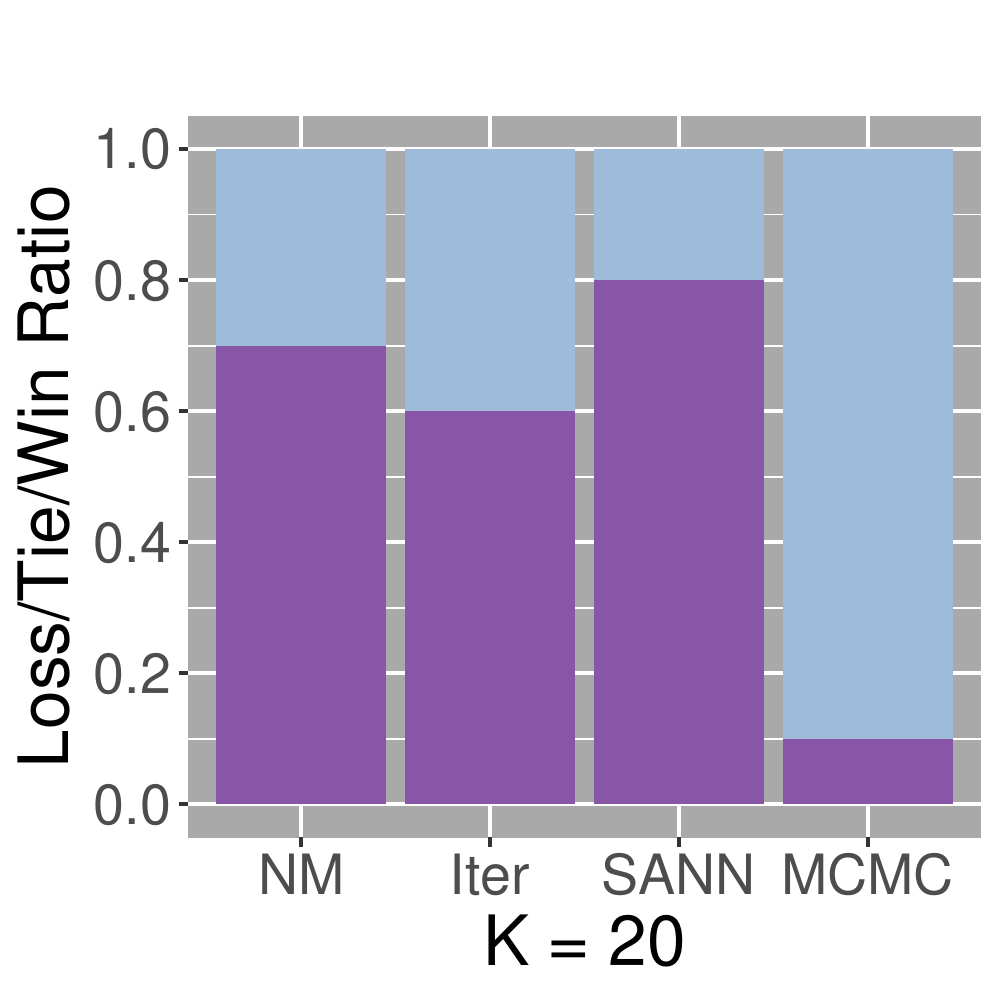} &
\includegraphics[scale=0.3]{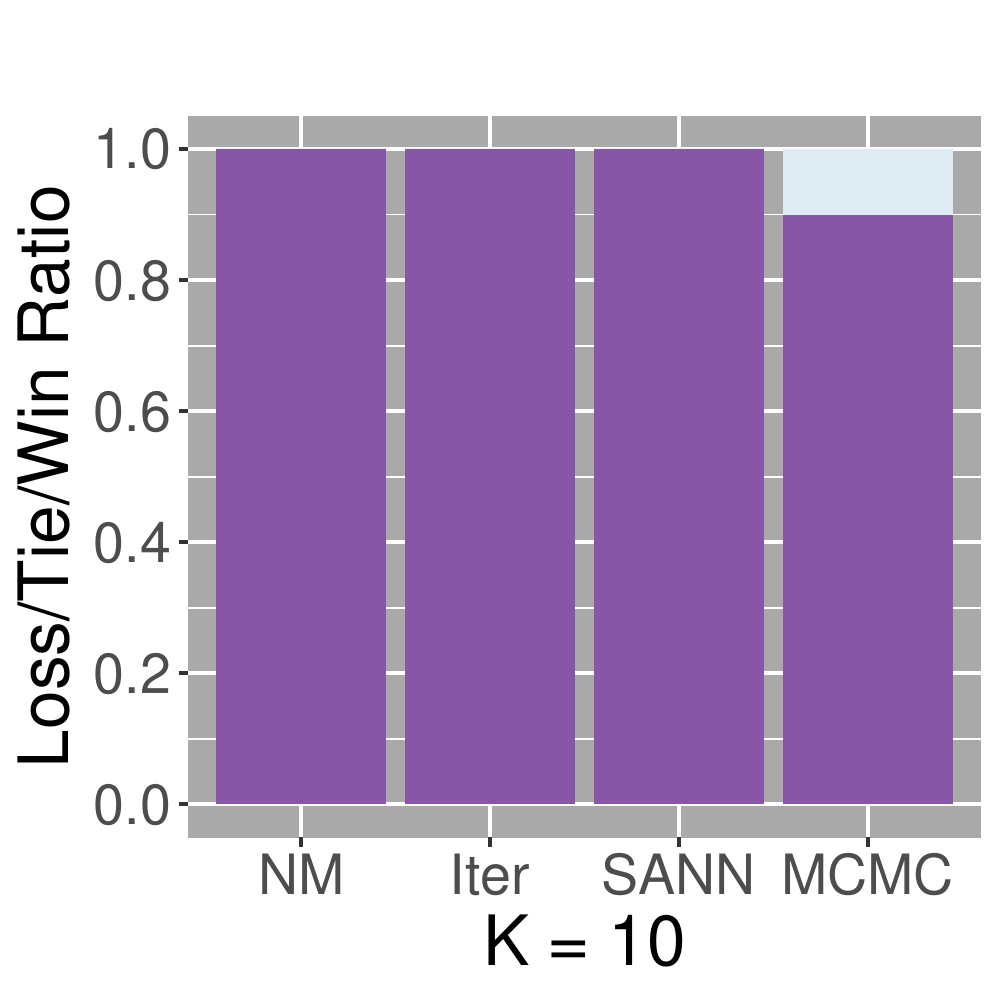}
\includegraphics[scale=0.3]{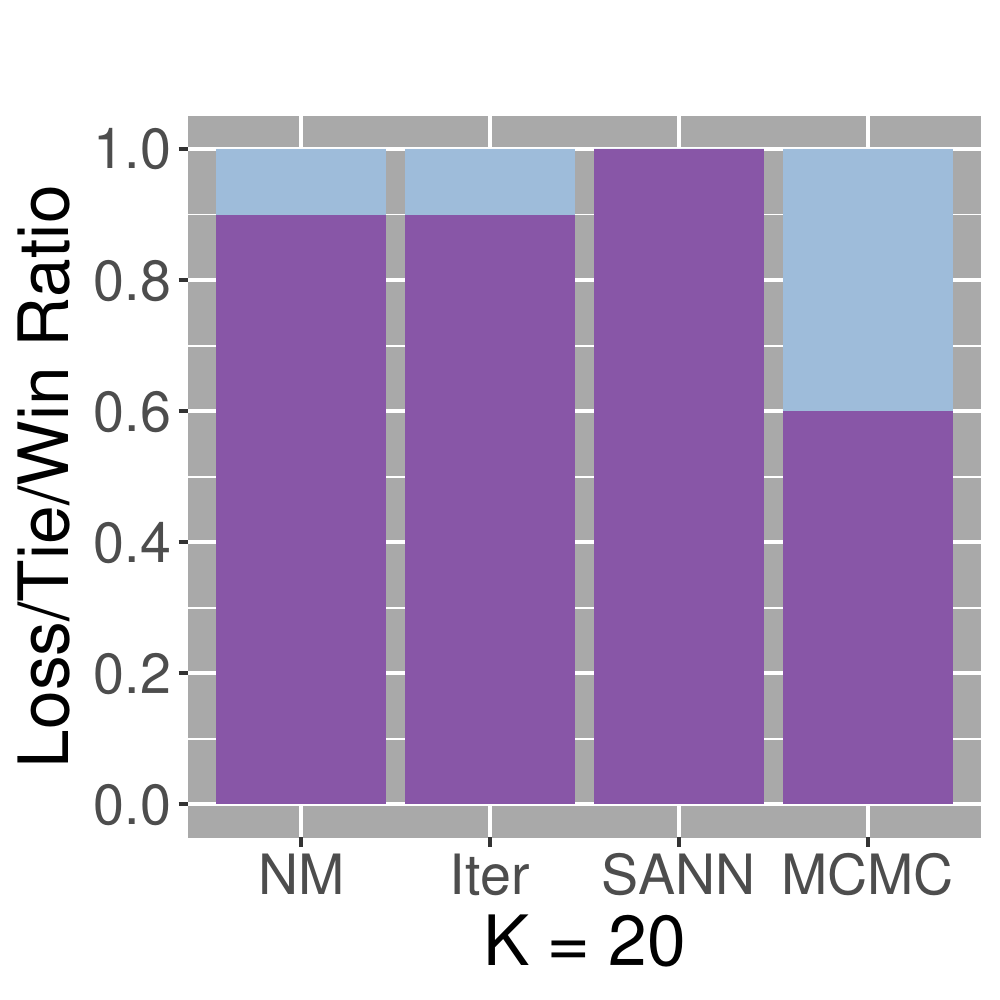}\\
\multicolumn{2}{c}{\includegraphics[scale=0.85]{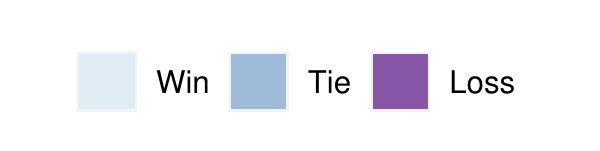}}
\end{tabular}
\end{figure}

We consider the following two regression models: for $i=1,\ldots, n$,
\eq{
  &  \mbox{Binary Regression: }    \hskip14pt y_i = 1\{x_i'\bt + \eps_i > 0\} \label{eq:binary}\\
  & \mbox{Censored Regression: } \hskip3pt y_i = \max\{x_i'\bt + \eps_i, 0\}
}
where $x_i$ is a $k$-dimensional vector generated from $N(0,I_k)$, $\eps_i$ is an error term generated from $N(0,0.25^2)$, and $\bt$ is a parameter of interest. The true parameter value is set to be $\bt_0=(1,\ldots,1)$. Recall that we do not know the true transformation function (binary or censored) of the data generating process when we estimate $\bt$ by the MRC estimator. For identification, we normalize the first coefficient of $\bt$ to be 1. We compare the performance of the MIP algorithm with that of the Nelder-Mead algorithm (NM), the iterative grid search (Iter-Grid), the simulated annealing (SANN), and the Markov Chain Monte Carlo method (MCMC). For all methods, the parameter space is set to be $\mathcal{B}=[-10,10]^{k-1}$. The time budget is set to be 600 seconds. The Nelder-Mead algorithm with repeated random starting points (Nelder-Mead 2 in the previous section) does not perform better (especially for large $k$) than Nelder-Mead 1 and is dropped in these simulation studies.

We first consider small-scale designs and check if the MIP algorithm achieves the global objective function. We set the sample size and the dimension of regressors to be $n=(50,100)$ and $k=(2,3)$, respectively for these small-scale designs. We next extend them into $n=(200,400)$ and $k=(10,20)$ and check the performance of the MIP algorithm in the limited time budget. Therefore, we consider 8 different designs in total in each regression model (binary/censored). We conduct 10 replications of each simulation design.

\begin{figure}[hp]
\caption{Loss/Tie/Win Ratio (Censored)} \label{fg:win-cen}
\centering
\begin{tabular}{cc}
$\underline{n=50}$ & $\underline{n=100}$ \\
\includegraphics[scale=0.3]{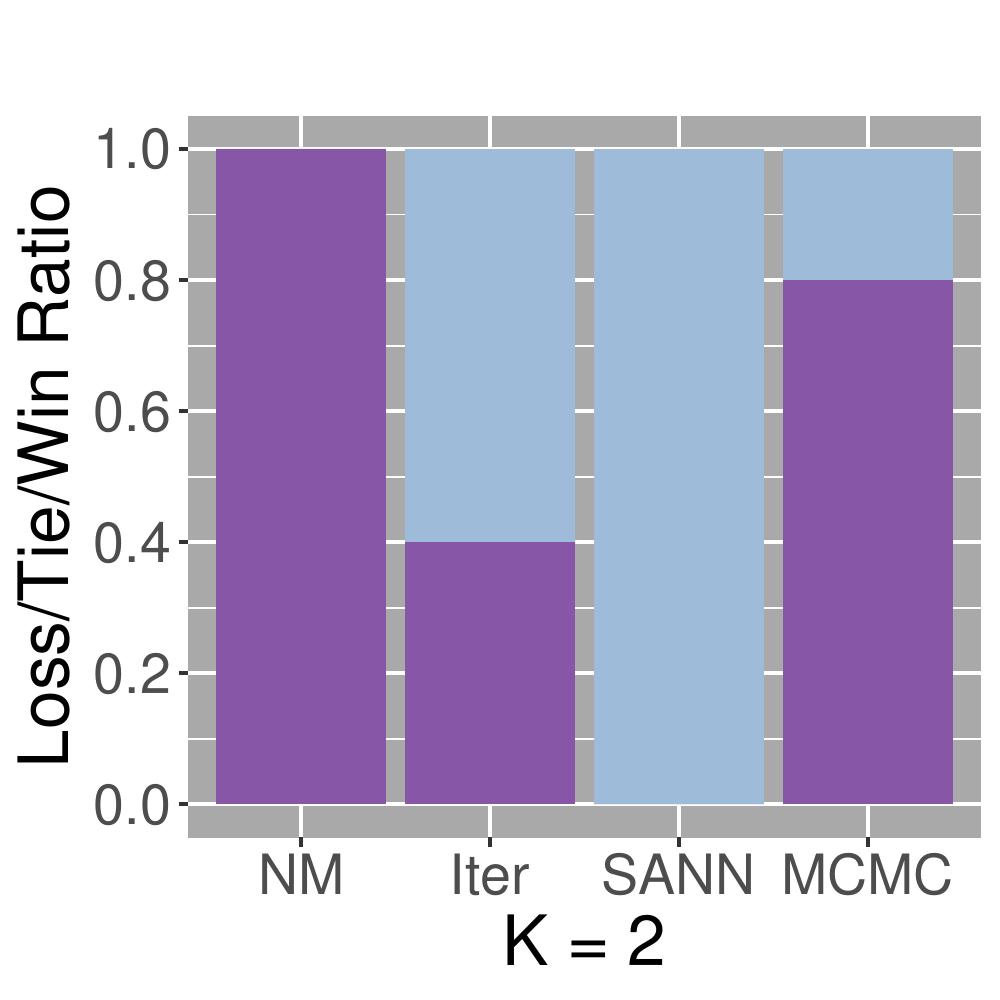}
\includegraphics[scale=0.3]{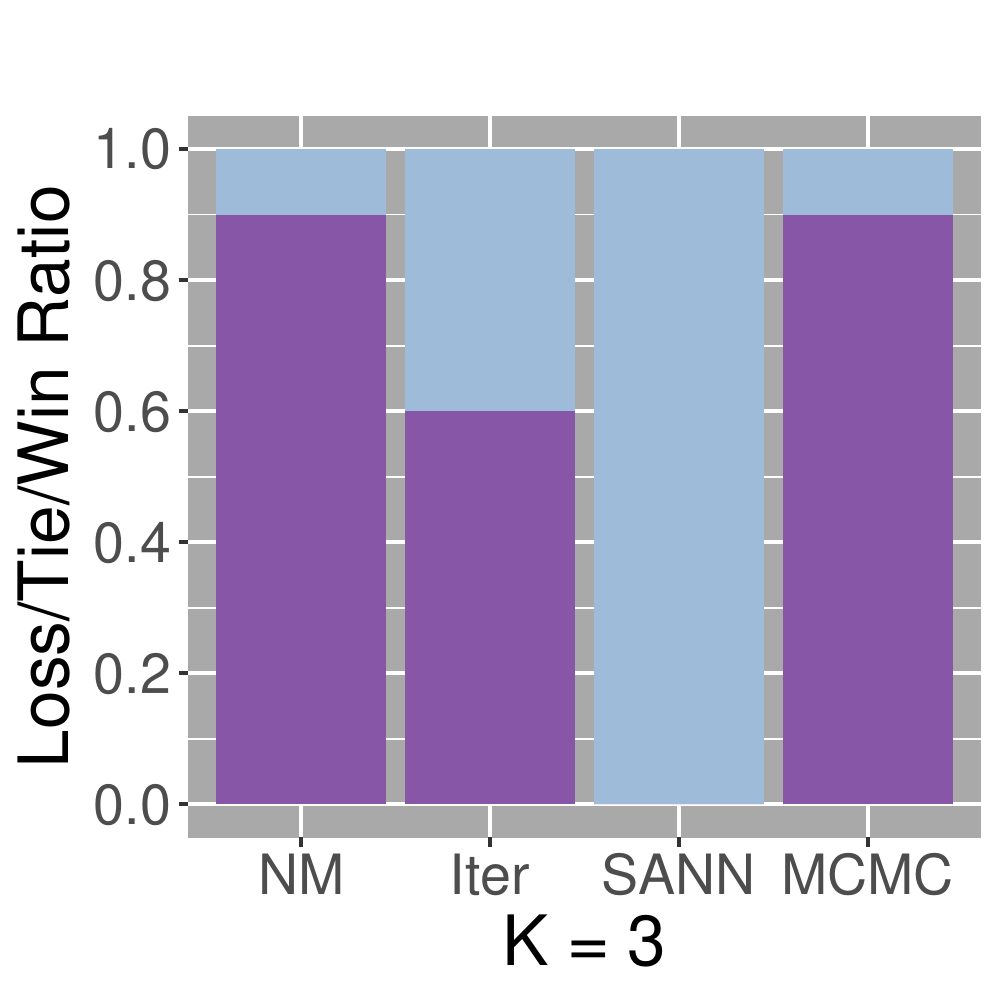} &
\includegraphics[scale=0.3]{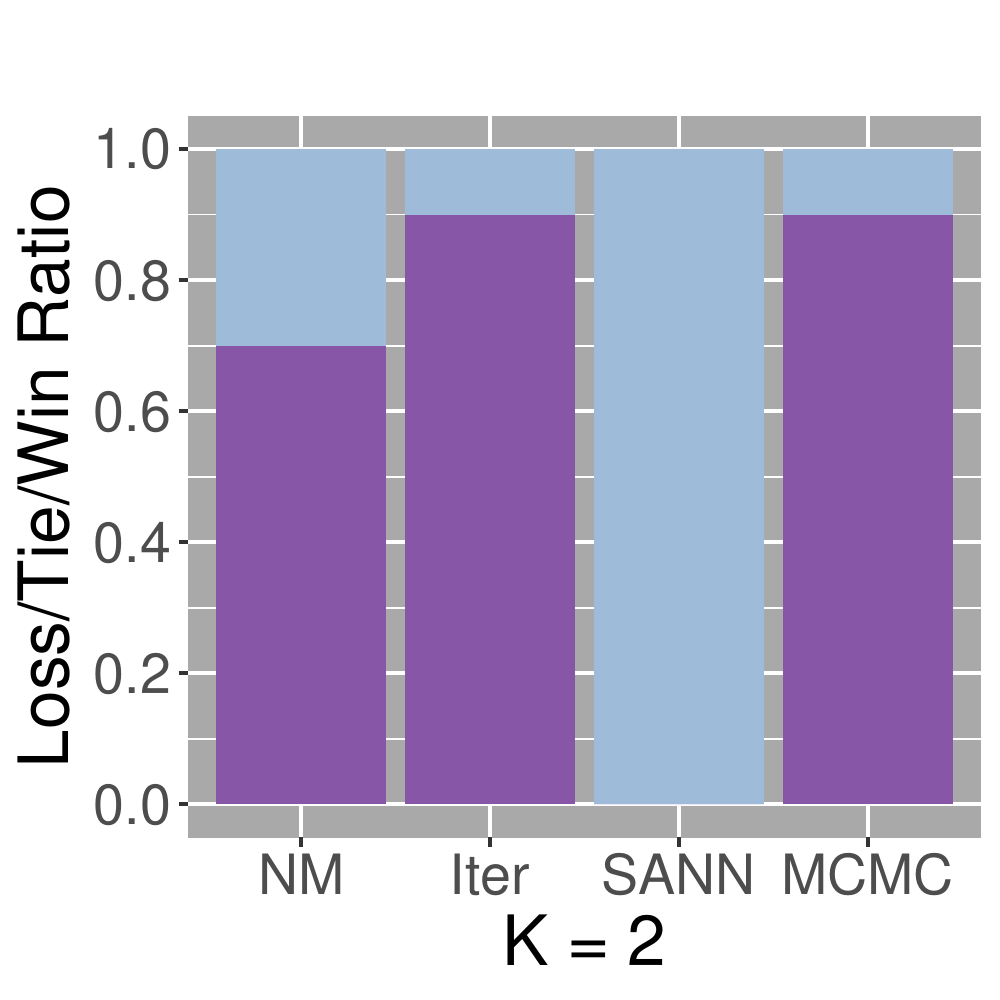}
\includegraphics[scale=0.3]{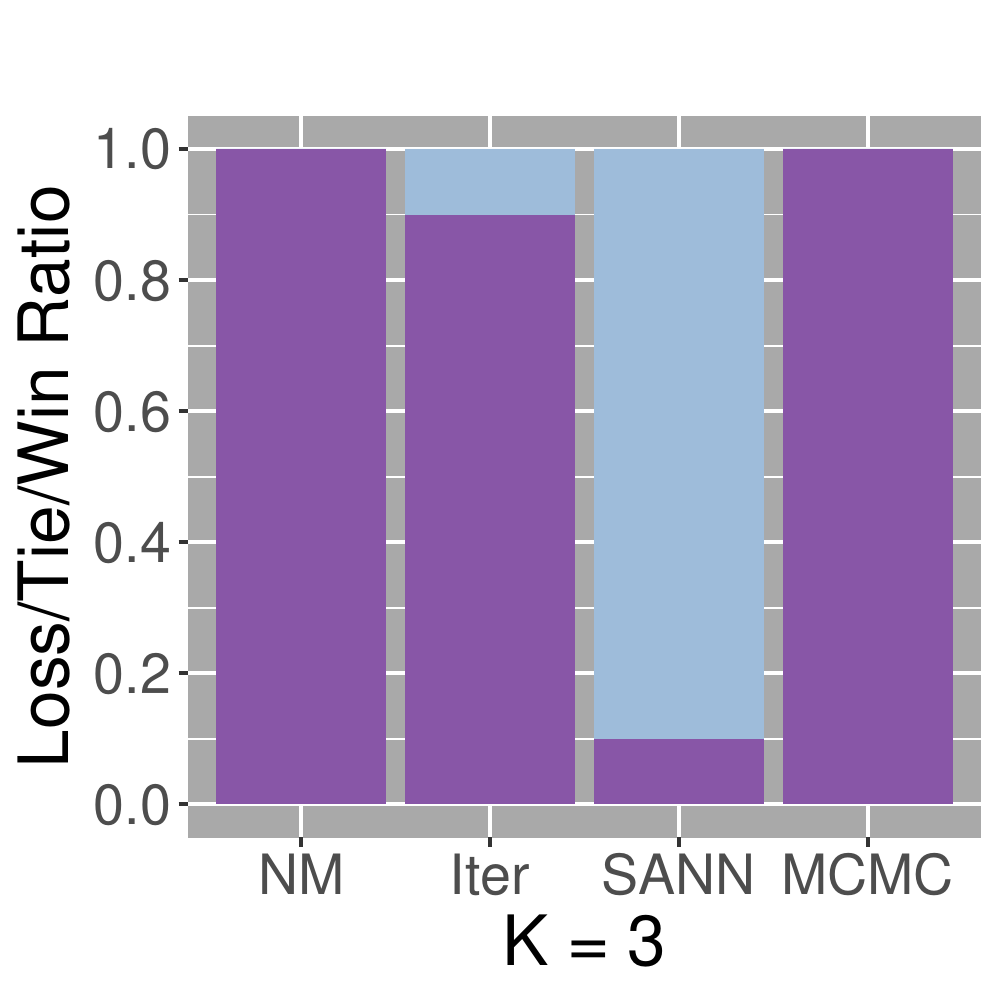}\\
$\underline{n=200}$ & $\underline{n=400}$ \\
\includegraphics[scale=0.3]{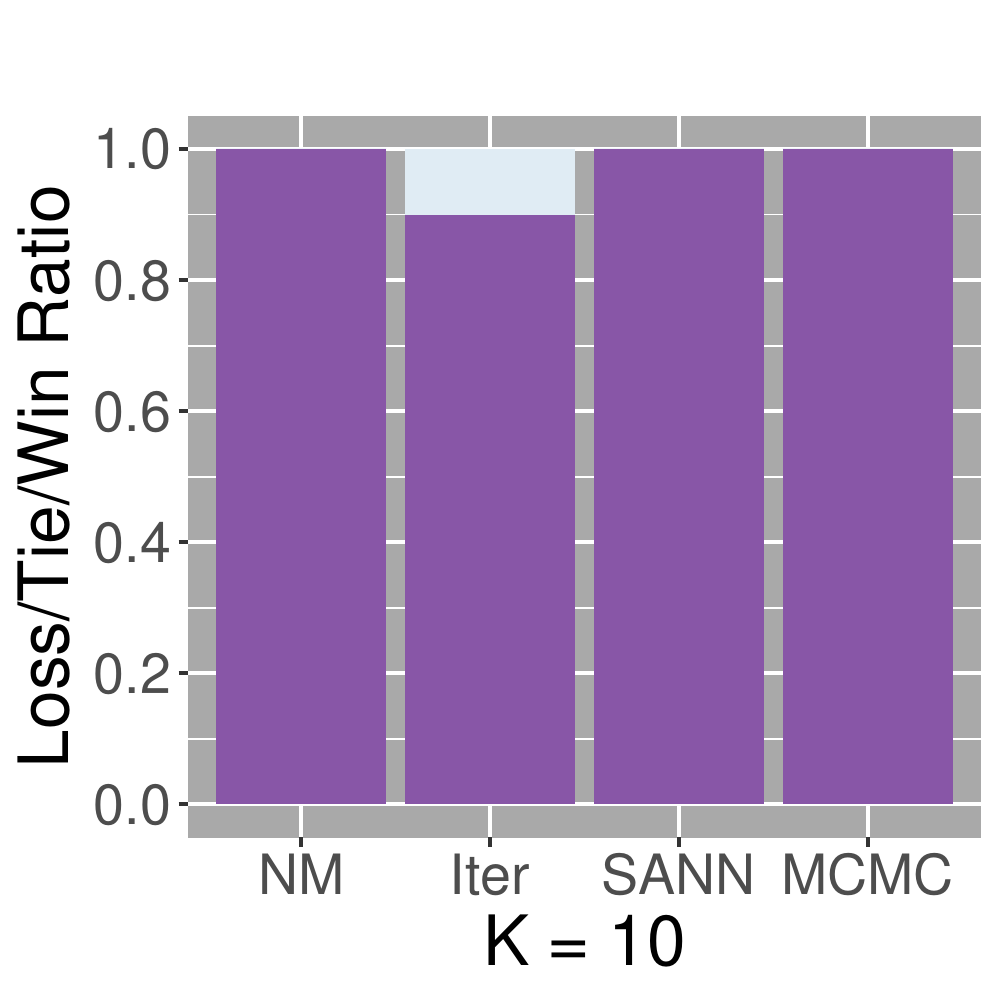}
\includegraphics[scale=0.3]{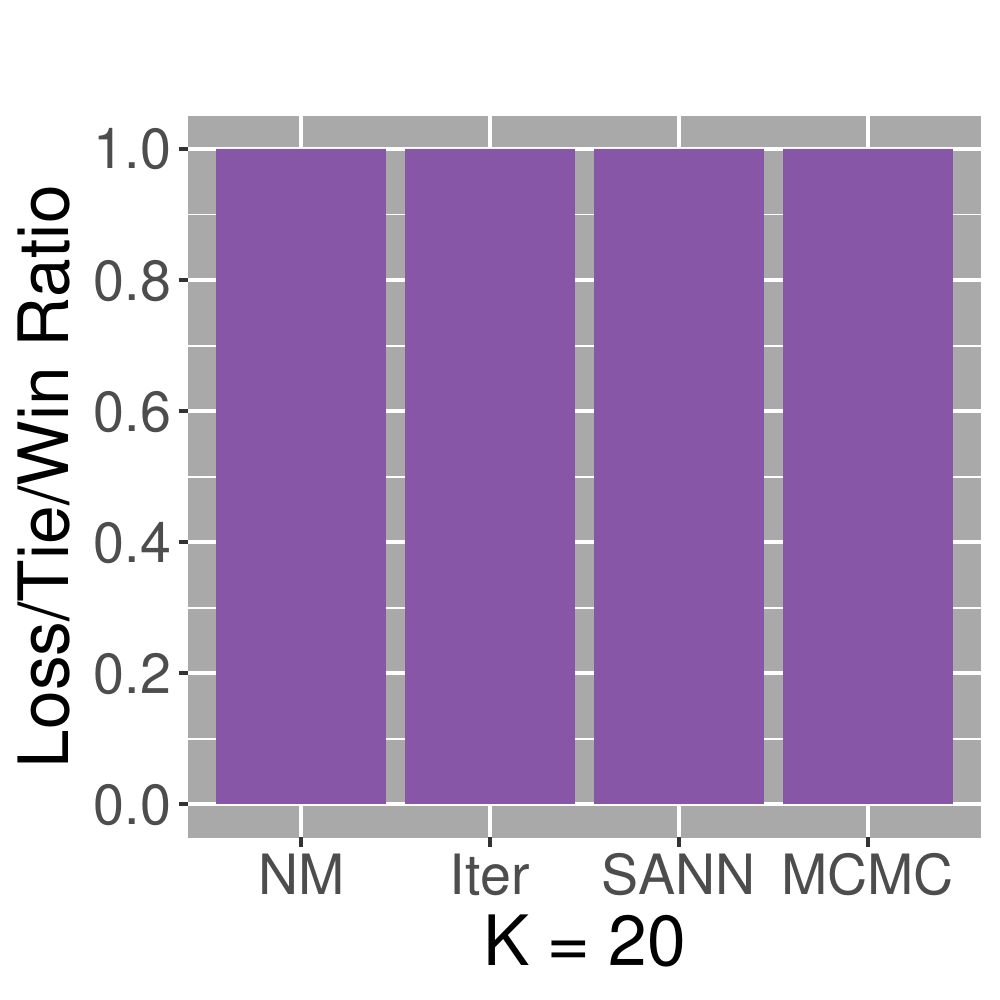} &
\includegraphics[scale=0.3]{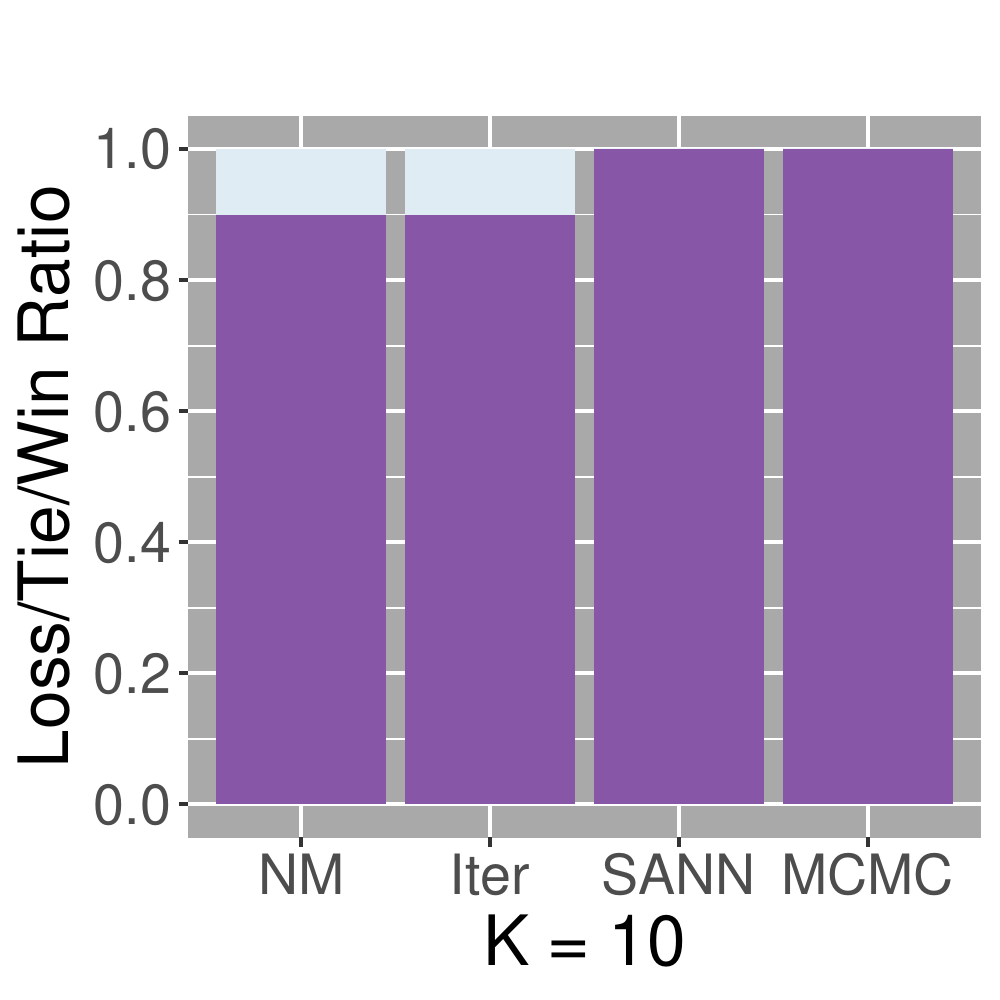}
\includegraphics[scale=0.3]{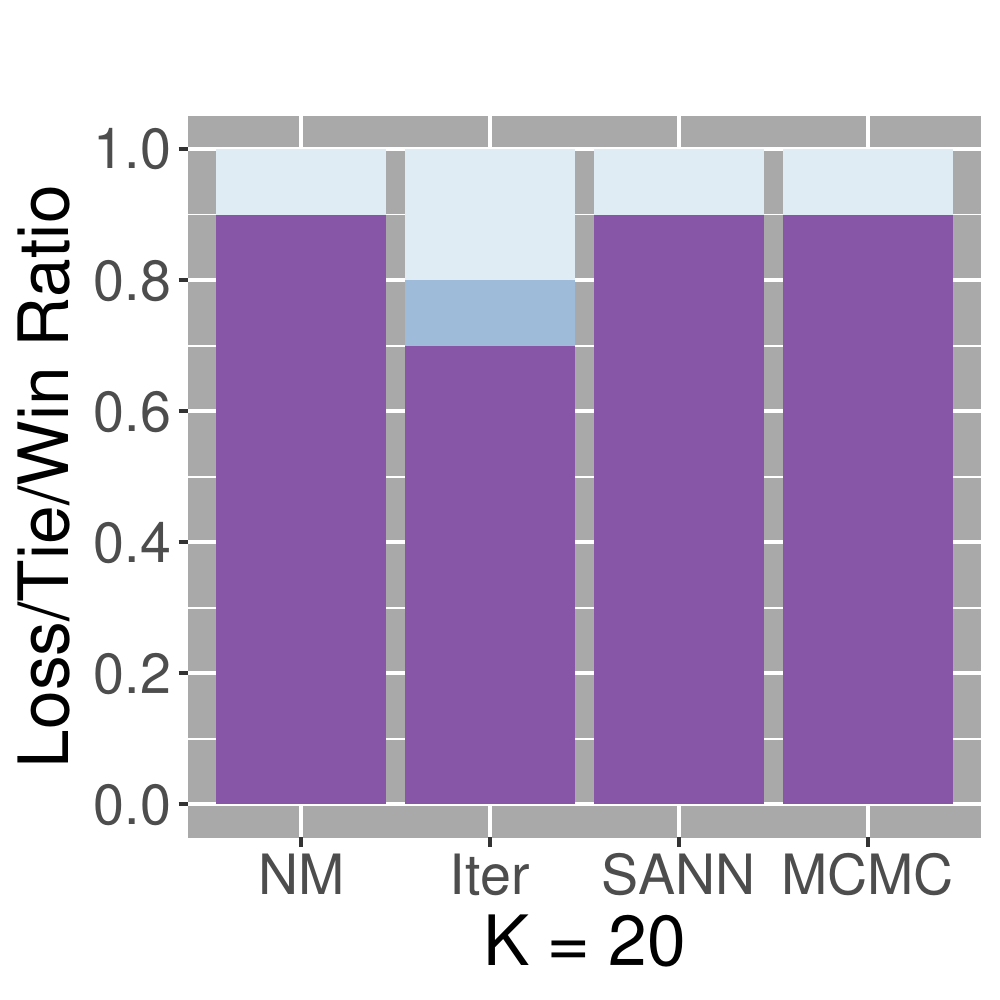}\\
\multicolumn{2}{c}{\includegraphics[scale=0.85]{figures/legend.pdf}}
\end{tabular}
\end{figure}

Figures \ref{fg:win-bin}--\ref{fg:win-cen} report the Loss/Tie/Win ratios of each alternative algorithm against MIP. In the graph, `Loss' means the objective function value of the algorithm is lower than that of MIP. `Tie' and `Win' are defined similarly. Overall, MIP outperforms the altenative methods. In the case of Binary Regression in Figure \ref{fg:win-bin}, MIP always achieves an equal or better objective function value than the alternative methods in all designs except one draw in $n=400, k=10$. In the small-scale design ($n=50, 100$ and $k=2,3$), SANN performs similar to MIP but the performance of MIP dominates in the large-scale design ($n=200,400$, $k=10,20$).
It is interesting that MIP performs better as the dimension of $k$ increases when $n=400$. As we confirm in Table \ref{tb:bin_time} below, MIP finds a more precise solution (lower MIP gap) in a substantially shorter time when $n=400$ and $k=20$ than when $n=400$ and $k=10$.

We observe similar patterns in Censored Regression in Figure \ref{fg:win-cen}.
The outperfomance of MIP is clearer in the large-scale designs ($n=200,400$ and $k=10,20$).
The overall performance of MIP in Censored Regression is better than that in Binary Regression when $n=200$. However, when $n=400$, MIP finds a worse solution than its competitors about 10-20\%. This is because the implied parameter space of $d_{ij}$ has a  much bigger dimension in Censored Regression than Binary Regression as it has less tied pairs of $(y_i,y_j)$. Recall that $d_{ij}$ is multiplied by 0 in the objective function when $y_i$ and $y_j$ are tied and we do not need to estimate such a $d_{ij}$.

\begin{table}[hp]
\begin{center}
\caption{Computation Time and MIP Gap (Binary)} \label{tb:bin_time}
\begin{tabular}{lrrrrr}
  \hline
& MIP & NM & Iter & SANN & MCMC \\
  \hline
 \multicolumn{6}{c}{\underline{$n=50, k=2$}}\\
 Max & 0.08 (0.00) & 0.19 & 0.12 & 0.80 & 0.59 \\
 Median & 0.04 (0.00) & 0.00 & 0.07 & 0.79 & 0.45 \\
 \\
  \multicolumn{6}{c}{\underline{$n=50, k=3$}}\\
 Max & 0.09 (0.00) & 0.00 & 0.26 & 0.97 & 0.51 \\
 Median & 0.07 (0.00) & 0.00 & 0.17 & 0.84 & 0.47 \\
 \\
\multicolumn{6}{c}{\underline{$n=100, k=2$}}\\
 Max & 0.47 (0.00) & 0.00 & 0.13 & 2.11 & 0.71 \\
 Median & 0.26 (0.00) & 0.00 & 0.13 & 2.06 & 0.69 \\
 \\
\multicolumn{6}{c}{\underline{$n=100, k=3$}}\\
 Max & 18.85 (0.00) & 0.00 & 0.66 & 2.23 & 0.75 \\
 Median & 0.64 (0.00) & 0.00 & 0.40 & 2.16 & 0.71 \\
 \\
\multicolumn{6}{c}{\underline{$n=200, k=10$}}\\
 Max & 28.81 (0.00) & 0.08 & 14.86 & 17.80 & 4.45 \\
 Median & 0.27 (0.00) & 0.05 & 14.29 & 17.27 & 4.24 \\
 \\
\multicolumn{6}{c}{\underline{$n=200, k=20$}}\\
 Max & 0.50 (0.00)    & 0.14 & 67.11 & 22.78 & 6.13 \\
 Median & 0.37 (0.00) & 0.12 & 34.18 & 22.02 & 5.68 \\
 \\
\multicolumn{6}{c}{\underline{$n=400, k=10$}}\\
 Max & 607.64 (0.29)    & 0.32 & 77.06 & 60.26 & 13.89 \\
 Median & 600.53 (0.09) & 0.20 & 49.24 & 58.55 & 13.16 \\
 \\
\multicolumn{6}{c}{\underline{$n=400, k=20$}}\\
 Max & 1.87 (0.00)    & 0.70 & 326.98 & 92.20 & 22.00 \\
 Median & 1.53 (0.00) & 0.54 & 211.61 & 91.18 & 21.25 \\
\hline
\end{tabular}
\end{center}
\renewcommand{\baselineskip}{11pt}
\textbf{Note:} MIP denotes the mixed integer programming method. NM does the Nelder-Mead simplex methods, Iter-Grid does the iterative grid search method, SANN does the simulated annealing method, and MCMC does the Markov Chain Monte Carlo method. The MIP gaps are given in the parentheses under the MIP column. The units for Time and MIP Gap are seconds and percent, respectively.
\end{table}

\begin{table}[hp]
\begin{center}
\caption{Computation Time and MIP Gap (Censored)} \label{tb:cen_time}
\begin{tabular}{lrrrrr}
 & MIP & NM & Iter-Grid & SANN & MCMC \\
  \hline
 \multicolumn{6}{c}{\underline{$n=50, k=2$}}\\
 Max    & 0.21 (0.00) & 0.00 & 0.10 & 1.14 & 0.51 \\
 Median & 0.15 (0.00) & 0.00 & 0.09 & 1.04 & 0.49 \\
 \\
 \multicolumn{6}{c}{\underline{$n=50, k=3$}}\\
 Max    & 8.80 (0.00) & 0.01 & 0.53 & 1.30 & 0.62 \\
 Median & 2.78 (0.00) & 0.00 & 0.28 & 1.03 & 0.51 \\
 \\
 \multicolumn{6}{c}{\underline{$n=100, k=2$}}\\
 Max & 20.18 (0.00)   & 0.00 & 0.17 & 3.11 & 1.00 \\
 Median & 5.82 (0.00) & 0.00 & 0.17 & 2.91 & 0.88 \\
 \\
 \multicolumn{6}{c}{\underline{$n=100, k=3$}}\\
 Max & 600.26 (0.34) & 0.01 & 2.08 & 6.94 & 2.22 \\
 Median & 317.32 (0.00) & 0.01 & 1.19 & 6.59 & 1.83 \\
 \\
 \multicolumn{6}{c}{\underline{$n=200, k=10$}}\\
 Max & 600.26 (2.13)    & 0.24 & 34.38 & 25.24 & 6.17 \\
 Median & 600.24 (1.21) & 0.09 & 21.98 & 23.79 & 5.63 \\
 \\
 \multicolumn{6}{c}{\underline{$n=200, k=20$}}\\
 Max  & 602.52 (1.11)   & 0.29 & 110.4 & 32.48 & 8.28 \\
 Median & 600.31 (0.88) & 0.21 & 79.79 & 31.23 & 7.61 \\
 \\
 \multicolumn{6}{c}{\underline{$n=400, k=10$}}\\
 Max  & 623.87 (1.99)   & 0.55 & 114.97 & 103.47 & 22.15 \\
 Median & 606.45 (1.72) & 0.40 & 88.23 & 93.78 & 20.69 \\
 \\
 \multicolumn{6}{c}{\underline{$n=400, k=20$}}\\
 Max & 610.85 (6.04)    & 1.35 & 490.84 & 167.22 & 38.82 \\
 Median & 602.57 (1.16) & 0.98 & 331.36 & 152.99 & 36.03 \\
   \hline
\end{tabular}
\end{center}
\renewcommand{\baselineskip}{11pt}
\textbf{Note:} See the note under Table \ref{tb:bin_time} for details.
\end{table}

Tables \ref{tb:bin_time}--\ref{tb:cen_time} provide some summary statistics of the computation time and the MIP gap. We first discuss the result of Binary Regression in Table \ref{tb:bin_time}. In small-scale designs, MIP requires about the same computation time as the alternative algorithms and it finds the global solution in less than a second except $n=100$ and $k=3$. In large-scale designs MIP is still able to find the global solution within the allocated time budget of 600 seconds, except when $n=400$ and $k=10$. In that design MIP hits the time limit of 600 seconds more often and the MIP gap does not achieve 0\%, i.e.\ we are not sure whether the solution is global or not. However, the gap size is quite small and less than 1\%. It is noteworthy that MIP performs much better in terms of the MIP gap when $k$ is bigger in large-scale designs. The computation time is even dramatically reduced when $n=400$.

We turn our attention to the result of Censored Regression in Table \ref{tb:cen_time}. As we discussed above, Censored Regression requires more computation time than Binary Regression and it mostly reaches the time limit of 600 seconds when $n=100$ and $k=3$. In large-scale designs, we observe the MIP gaps larger than 1\%, which could be the reason that the solutions of MIP are sometimes worse than those of the alternative algorithms. Other patters are quite similar to those in Binary Regression including that the performance of MIP becomes better when $k$ is higher for large-scale designs.

In sum, the performance of the proposed MIP algorithm for the MRC estimator is satisfactory. It always finds the global solution in small-scale designs where the existing methods fail to do quite often. Furthermore, it performs better than the alternative algorithms even in large-scale designs by spending a feasible amount of computation time. The MIP gap also provides useful guidance for the quality of a solution in hand when a researcher should stop searching for the global solution because of the time limit.

%
%

\section{Conclusion}

In this paper we propose a feasible computation method to get a global solution for the maximum rank correlation estimator of \cite{han1987non} using the mixed integer programming (MIP).
We show that the proposed MIP method outperforms the alternative methods in the empirical example of female labor-force participation.
One advantage of the proposed MIP method is that it can be easily extended to many constraint rank maximization problems as illustrated in the best subset rank prediction problem, where we also prove that the non-asymptotic bound of the tail probability decays exponentially. This result sheds light on the research of the high-dimensional rank estimation models, which we leave for future research.


    \bibliographystyle{chicago}
        {
        \bibliography{ST-mio}
        }

    \clearpage

    \section*{Appendix A: Proof of Theorem 3.1}
    \renewcommand{\theequation}{A.\arabic{equation}}
    \renewcommand{\thelemma}{A.\arabic{lemma}}
    \renewcommand{\thesection}{A}
    \setcounter{equation}{0}

    \medskip

    In this appendix we provide the proof of Theorem \ref{thm-main}. We first prove some useful lemmas. We need the following notation. Let $m$ be a subset of the index set $\{1,2,\ldots, p\}$, where $\vert m \vert =s$. Let $\mathcal{M}$ be the collection of $m$. Thus, for any given $p$ and $s$, $\lt\vert \mathcal{M} \rt\vert = \binom{p}{s}$. Let $\mathcal{B}_m := \{ \bt \in \mathcal{B}: \bt_j =0 \mbox{ if  }j \notin m\}$.
Let $w:=(y,x')$ and $\mathcal{W}$ be the support of $w$. Define $f_{\bt}: \mathcal{W}\times \mathcal{W} \mapsto \{0,1\}$ as
\eqs{
    f_{\bt}(w_i,w_j) & := 1\{ 1\{y_i> y_j \} = R_{\bt}(x_i,x_j) \} \\
                             & =  1 - 1\{y_i > y_j\} + (2\cdot 1\{y_i > y_j\} - 1 ) R_{\bt}(x_i,x_j).
}Let $\mathcal{F}_m:=\{f_{\bt}(\cdot,\cdot): \bt \in \mathcal{B}_m \}$.

From the Hoeffding decomposition, we have
\eq{
    S_n(\bt) = S(\bt) +\frac{1}{n} \sum_{i=1}^n g_{\bt}(w_i) + \frac{2}{n(n-1)} \sum_{i=1}^{n} \sum_{j \neq i} h_{\bt} (w_i,w_j), \label{eq:hoeffding}
}
where
\eq{
    g_{\bt}(w_i) & := \int_{\mathcal{W}} f_{\bt} (w_i,w) dP(w) + \int_{\mathcal{W}} f_{\bt} (w,w_i) dP(w) -2 S(\bt) \label{eq:g-function}\\
    h_{\bt}(w_i,w_j) & := f_{\bt}(w_i,w_j) - \int_{\mathcal{W}} f_{\bt} (w_i,w) dP(w) - \int_{\mathcal{W}} f_{\bt} (w,w_j) dP(w) + S(\bt). \label{eq:h-function}
}Note that the last term of \eqref{eq:hoeffding} is a $P$-degenerate U-process.

    \begin{lemma}\label{lm:h-bound}
For the measurable function $h_{\bt}(\cdot,\cdot)$ defined in \eqref{eq:h-function}, the following inequality holds for some universal constants $C_k>0$, $k=1,\ldots,5$:
\eq{
\Pr \lt( \sup_{\bt \in \mathcal{B}_m} \lt\vert \frac{2}{n(n-1)}\sum_{i=1}^n \sum_{j\neq i} 2^{-1}h_{\bt}(w_i, w_j) > \frac{t}{{n}} \rt) \rt\vert \le C_1 (512eC_4)^{(16s+16)}e^{- C_2 t}, \label{eq:h-uniform}
}if $C_5(16s+16+\gm)^{3/2}\log 2 \le t \le n$, where $\gm:=\max(\log(512 e C_4)/\log n, 0)$.
\end{lemma}

    \textbf{Proof of Lemma \ref{lm:h-bound}:}

    From Lemma 9.6, 9.9(iii), 9.9(vi), and 9.9(v) in \citet{kosorok2007introduction}, the VC-index of $\mathcal{F}_m$, $V(\mathcal{F}_m)$, satisfies that
\eq{
V(\mathcal{F}_m)\le 2s+3. \label{eq:VC-index-F}
}We now define a pseudometric for two measurable functions $f$ and $g$ in $\mathcal{F}_m$:
\eqs{
    d_Q(f,g) := \lt( \int (f-g)^2 dQ  \rt)^{1/2}.
}Let $\eps>0$ be given. Then, the covering number $N(\eps,\mathcal{F}_m,d_Q)$ is defined as the minimal number of
open $\eps$-balls required to cover $\mathcal{F}_m$. Noting that $\mathcal{F}_m$ has a constant envelope $f=1$, we apply Theorem 9.3 in \citet{kosorok2007introduction} to get
\eq{\label{eq:N-bound}
\begin{split}
    \sup_Q N(\eps, \mathcal{F}_m,d_Q) & \le C_3 (2s+3) (4e)^{2s+3} \lt(\frac{2}{\eps}\rt)^{4(s+1)} \\
    & \le \lt(\frac{(C_3(2s+3)(16e)^{2s+3})^{1/(4s+4)}}{\eps}\rt)^{4s+4} \\
    & \le \lt(\frac{32 e C_4}{\eps}\rt)^{4s+4},
\end{split}
}where $C_4:=\max(1,C_3)$ is a universal constant. The last inequality comes from
\eqs{
(C_3(2s+3)(16e)^{2s+3})^{1/(4s+4)} \le 16 e C_4 (2s+3)^{1/(4s+4)} \le 16 e C_4 2^{(2s+3)/(4s+4)} \le 32 e C_4.
}

We now define the following classes of functions:
\eqs{
    \mathcal{F}_{m,1} & :=\lt\{\int_{\mathcal{W}} f_{\bt}(\cdot,w) dP(w): f_{\bt} \in \mathcal{F}_m \rt\} \\
    \mathcal{F}_{m,2} & :=\lt\{\int_{\mathcal{W}} f_{\bt}(w,\cdot) dP(w): f_{\bt} \in \mathcal{F}_m \rt\}
}Define a pseudometric for the functions $f$ and $g$ in $\mathcal{F}_{m,1}$, $\mathcal{F}_{m,1}$ as
\eqs{
    d_P(f,g) := \lt( \int (f-g)^2 dP  \rt)^{1/2}.
}From Lemma 20 in \citet{nolan1987u}, we have
\eq{
    \sup_P N(\eps, \mathcal{F}_{m,j}, d_P ) \le \sup_Q N(\eps, \mathcal{F}_{m}, d_Q ) \le \lt(\frac{32C_3e}{\eps}\rt)^{4s+4} \label{eq:F_m_j-covering-bound}
}for $j=1,2$. Using the same arguments, we have
\eq{
    N(\eps, S(\bt), d_2 ) \le \sup_P N(\eps, \mathcal{F}_{m,1}, d_P ) \le \lt(\frac{32C_3e}{\eps}\rt)^{4s+4}  \label{eq:S-covering-bound}
}where $d_2 := \lt( \int (f-g)^2 dw \rt)^{1/2}$ for $f,g \in S(\bt)$.

We now consider the class of $P$-degenerate functions $h_{\bt}(\cdot,\cdot)$ defined in \eqref{eq:h-function}:
\eqs{
    \mathcal{H}_m := \lt\{2^{-1} h_{\bt}(\cdot,\cdot): \bt \in \mathcal{B}_m  \rt\}.
}Using the results in \eqref{eq:N-bound}, \eqref{eq:F_m_j-covering-bound}, \eqref{eq:S-covering-bound}, and Lemma 16 in \citet{nolan1987u}, we get
\eqs{
    \sup_{Q} N(\eps, \mathcal{H}_m, d_Q)
    \le \lt( \frac{512 e C_4 }{\eps} \rt)^{16s+16}.
}The desired result is established by applying Theorem 6.3 in \citet{major2005tail}.

    \hfill$\square$\\

    \begin{lemma}\label{lm:g-bound}
For the measurable function $h_{\bt}(\cdot,\cdot)$ defined in \eqref{eq:h-function}, the following inequality holds for some universal constants $D_1>0$:
\eq{
\Pr \lt( \sup_{\bt \in \mathcal{B}_m} \lt\vert \frac{1}{n}\sum_{i=1}^n 2^{-1}g_{\bt}(w_i) \rt\vert > \frac{t}{\sqrt{n}} \rt)  \le \lt( \frac{D_1 t}{\sqrt{12s+12}}\rt)^{12s+12}e^{- 2t^2}. \label{eq:g-uniform}
}
\end{lemma}
    \textbf{Proof of Lemma \ref{lm:g-bound}:}
    Let $\mathcal{G}_m:=\{2^{-1}g_{\bt}(\cdot): \bt \in \mathcal{B}_m\}$. From \eqref{eq:F_m_j-covering-bound}, \eqref{eq:S-covering-bound}, and Lemma 16 in \citet{nolan1987u}, we have
\eq{
N(\eps, \mathcal{G}_m, d_P) \le \lt(  \frac{256e C_5 }{\eps} \rt)^{12s+12}.
}for a universal constant $C_5>0$. Then, the desired result is established by applying Theorem 1.3 in \citet{talagrand1994sharper}
    \hfill$\square$\\

We now ready to prove the main theorem.
Using $S(\bt)\ge0$, $S_n(\bt)\ge0$, the triangular inequality, and $S_n(\what{\bt}) \ge S(\what{\bt})$, we have
\eq{
    U_n & = \sup_{\bt\in\mathcal{B}_n} S(\bt) - S(\what{\bt}) \notag \\
        & = \sup_{\bt\in\mathcal{B}_n} \vert S(\bt) - S_n(\bt) + S_n(\bt) \vert - S(\what{\bt}) \notag \\
        & \le \sup_{\bt\in\mathcal{B}_n} \vert  S_n(\bt) - S(\bt) \vert  + \sup_{\bt\in\mathcal{B}_n}  S_n(\bt)  - S(\what{\bt}) \notag \\
        & =  \sup_{\bt\in\mathcal{B}_n} \vert  S_n(\bt) - S(\bt) \vert  + S_n(\what{\bt})  - S(\what{\bt}) \notag \\
        & \le 2 \sup_{\bt\in\mathcal{B}_n} \vert  S_n(\bt) - S(\bt) \vert. \label{eq:U_n-last-term}
}
Using \eqref{eq:U_n-last-term} and \eqref{eq:hoeffding}, we have

{
\footnotesize
\eq{
    P\lt(  U_n > 4\sqrt{\frac{M_{\sigma}r_n}{n}} \rt)
    & \le P\lt(  2 \sup_{\bt\in\mathcal{B}_s} \vert  S_n(\bt) - S(\bt) \vert > 4\sqrt{\frac{M_{\sigma}r_n}{n}} \rt) \notag \\
    & \le P\lt(   \sup_{\bt\in\mathcal{B}_s} \lt\vert \frac{1}{n}\sum_{i=1}^n g_{\bt}(w_i) + \frac{2}{n(n-1)} \sum_{i=1}^{n} \sum_{j>i}h_{\bt} (w_i,w_j) \rt\vert > 2\sqrt{\frac{M_{\sigma}r_n}{n}} \rt) \notag \\
    \begin{split}
    & \le P\lt(   \sup_{\bt\in\mathcal{B}_s} \lt\vert \frac{1}{n}\sum_{i=1}^n 2^{-1}g_{\bt}(w_i)  \rt\vert > \sqrt{\frac{M_{\sigma}r_n}{n}} \rt)  \label{eq:split-u-processes}\\
    & \hskip10pt + P \lt(   \sup_{\bt\in\mathcal{B}_s} \lt\vert  \frac{2}{n(n-1)} \sum_{i=1}^{n} \sum_{j>i} 2^{-1} h_{\bt} (w_i,w_j) \rt\vert > \sqrt{\frac{M_{\sigma}r_n}{n}} \rt).
    \end{split}
}
\par
}
We first calculate the upper bound of the first term in \eqref{eq:split-u-processes}. Let $t=M_{\sigma}r_n$. Using the definition of $\mathcal{B}_m$, $\vert \mathcal{M} \vert = \binom{p}{s} \le p^s$, and Lemma \ref{lm:g-bound}, we have
\eq{
P\lt(   \sup_{\bt\in\mathcal{B}_s} \lt\vert \frac{1}{n}\sum_{i=1}^n 2^{-1}g_{\bt}(w_i)  \rt\vert > \frac{t}{\sqrt{n}} \rt)
& \le \sum_{m\in \mathcal{M}} P\lt(   \sup_{\bt\in\mathcal{B}_m} \lt\vert \frac{1}{n}\sum_{i=1}^n 2^{-1}g_{\bt}(w_i)  \rt\vert > \frac{t}{\sqrt{n}}  \rt) \notag \\
& \le p^s \lt( \frac{D_1 t}{\sqrt{12s+12}}\rt)^{12s+12}e^{- 2t^2}, \label{eq:g-exp-bound}
}for a universal constant $D_1$. Let $D_{\sigma}:=D_{\sigma,1} \vee D_{\sigma,2}$, where $D_{\sigma,1}:=2^{-1}(1+\sigma) \vee D_1^2$ and $D_{\sigma,2}$ is a universal constant that will be defined later.  Since $t \ge D_1$ and $s \ge 1$, the bound in \eqref{eq:g-exp-bound} is further bounded as
\eqs{
p^s \lt( \frac{D_1 t}{\sqrt{12s+12}}\rt)^{12s+12}e^{- 2t^2} \le e^{\ld_1(s,p,t)},
}where $\ld_1(s,p,t):= -2t^2 + (24s+24)\ln t + s \ln p -(24s+24) \ln 2$. By the definition of $t$, $r_n\ge s\ln p$, condition \eqref{eq:thm1-con1}, and the definition of $D_{\sigma}$, we have
\eqs{
\ld_1(s,p,t)    & \le (-2D_{\sigma} + 1) r_n + (12s + 12) \ln(D_{\sigma} r_n) - (24s+24) \ln2 \\
            & \le (-2D_{\sigma} + 2) r_n \\
            & \le -\sigma r_n.
}Therefore, it is established that
\eq{
P\lt(   \sup_{\bt\in\mathcal{B}_s} \lt\vert \frac{1}{n}\sum_{i=1}^n 2^{-1}g_{\bt}(w_i)  \rt\vert > \sqrt{\frac{M_{\sigma}r_n}{n}} \rt)  \le e^{-\sigma r_n}. \label{eq:g-final-result}
}

We next turn our attention to the second term in \eqref{eq:split-u-processes}. Using the similar arguments in \eqref{eq:g-exp-bound} with Lemma \ref{lm:h-bound}, we have
\eq{
P \lt(   \sup_{\bt\in\mathcal{B}_s} \lt\vert  \frac{2}{n(n-1)} \sum_{i=1}^{n} \sum_{j>i} 2^{-1} h_{\bt} (w_i,w_j) \rt\vert > \frac{\sqrt{n} t}{n} \rt)
& \le \sum_{m\in \mathcal{M}} P\lt(   \sup_{\bt\in\mathcal{B}_m} \lt\vert \frac{1}{n}\sum_{i=1}^n 2^{-1}g_{\bt}(w_i)  \rt\vert > \frac{\sqrt{n} t}{n} \rt) \notag \\
& \le p^s C_1 (512eC_4)^{(16s+16)}e^{- C_2 \sqrt{n} t}, \label{eq:h-exp-bound}
}for some universal constants $C_k>0$, $k=1,2,4$. Let $D_{\sigma,2} := C_2^{-1}(2+\sigma) \vee (C_1^2 \vee (eC_4)^2)$. From $t \le \sqrt{n}$, $C_1 < t$, and $eC_4 < t$, the bound in \eqref{eq:h-exp-bound} is further bounded as
\eqs{
 p^s C_1 (512eC_4)^{(16s+16)}e^{- C_2 \sqrt{n} t} \le e^{\ld_2(s,p,t)},
}where $\ld_2(s,p,t):= -C_2 t^2 + (16s+17)\ln t + s \ln p + 9(16s+16) \ln 2$. Recall that $D_{\sigma} = D_{\sigma,1} \vee D_{\sigma,2}$. Using the similar arguments as above, we have
\eqs{
\ld_2(s,p,t)
& \le (-C_2 D_{\sigma} + 1) r_n + \lt(8s + \frac{17}{2} \rt) \ln (D_{\sigma} r_n) + 9(16s+16) \ln 2 \\
& \le (-C_2 D_{\sigma} + 2) r_n \\
& \le -\sigma r_n.
}Therefore, it is established that
\eq{
P \lt(   \sup_{\bt\in\mathcal{B}_s} \lt\vert  \frac{2}{n(n-1)} \sum_{i=1}^{n} \sum_{j>i} 2^{-1} h_{\bt} (w_i,w_j) \rt\vert > \frac{t}{\sqrt{n}} \rt) \le -\sigma r_n. \label{eq:h-final-result}
}
From \eqref{eq:split-u-processes}, \eqref{eq:g-final-result}, and \eqref{eq:h-final-result}, we establish the desired result:
\eqs{
P\lt(  U_n > 4\sqrt{\frac{M_{\sigma}r_n}{n}} \rt)  \le -2\sigma r_n.
}

\end{document}